\documentclass[usenatbib]{mnras}
\usepackage{newtxtext,newtxmath}
\usepackage{amsmath}
\usepackage{verbatim,ulem}
\usepackage{url}
\usepackage{fixltx2e} 
\usepackage{subcaption}
\usepackage{bm}
\usepackage{tikz}
\usepackage[export]{adjustbox}
\usetikzlibrary{shapes.misc,shadows}
\usepackage{ctable}
\usepackage{CJK}

\newcommand{\be}{\begin{equation}}
\newcommand{\ee}{\end{equation}}

\newcommand{\etal}{et al.}

\newcommand{\degree}{^{\circ}}

\newcommand{\Alf}{{Alfv\'en}}

\newcommand{\gizmourl}{\href{http://www.tapir.caltech.edu/~phopkins/Site/GIZMO.html}{\url{http://www.tapir.caltech.edu/~phopkins/Site/GIZMO.html}}}

\newcommand{\acknowledgments}{\begin{small}\section*{Acknowledgments}\end{small}}

\newcommand{\dataavailability}[1]{\begin{small}\section*{Data Availability}\end{small}{\noindent #1}\vspace{5pt}}

\definecolor{darkgreen}{RGB}{0,100,0}

\title[CR Confinement by Dust]{Numerical Study of Cosmic Ray Confinement through Dust Resonant Drag Instabilities}

\author[Ji \etal]{ \parbox[t]{\textwidth}{Suoqing Ji (季索清)$^{1,2}$, Jonathan
Squire$^3$ and Philip F.~Hopkins$^2$ }\vspace*{4pt} \\
$^1$ Astrophysics Division \& Key Laboratory for Research in Galaxies and Cosmology, Shanghai Astronomical Observatory, Chinese Academy of Sciences, \\
\,Shanghai 200030, China. E-mail:suoqing@shao.ac.cn \\
$^2$ TAPIR \& Walter Burke Institute for Theoretical Physics, Mailcode 350-17, California Institute of Technology, Pasadena, CA
91125, USA. \\
$^3$ Physics Department, University of Otago, 730 Cumberland St., Dunedin 9016,
New Zealand \\
}

\date{}
\begin{document}
\begin{CJK}{UTF8}{gbsn}
\maketitle
\label{firstpage}

\begin{abstract}
We investigate the possibility of cosmic ray (CR) confinement by charged dust
grains through resonant drag instabilities (RDIs). We perform
magnetohydrodynamic particle-in-cell simulations of magnetized gas mixed with
charged dust and cosmic rays, with the gyro-radii of dust and GeV CRs on
$\sim\mathrm{AU}$ scales fully resolved. As a first study, we focus on one type
of RDI wherein charged grains drift super-Alfv{\'e}nically, with Lorentz forces
strongly dominating over drag forces. Dust grains are unstable to the RDIs and
form concentrated columns and sheets, whose scale grows until saturating at the
simulation box size. Initially perfectly-streaming CRs are strongly scattered by
RDI-excited Alfv{\'e}n waves, with the growth rate of the CR perpendicular
velocity components equaling the growth rate of magnetic field perturbations.
These rates are well-predicted by analytic linear theory. CRs finally become
isotropized and drift at least at $\sim v_\mathrm{A}$ by unidirectional
Alfv\'{e}n waves excited by the RDIs, with a uniform distribution of the pitch
angle cosine $\mu$ and a flat profile of the CR pitch angle diffusion
coefficient $D_{\mu\mu}$ around $\mu = 0$, without the ``$90\degree$ pitch angle
problem.'' With CR feedback on the gas included, $D_{\mu\mu}$ decreases by a
factor of a few, indicating a lower CR scattering rate, because the backreaction
on the RDI from the CR pressure adds extra wave damping, leading to lower
quasi-steady-state scattering rates. Our study demonstrates that the
dust-induced CR confinement can be very important under certain conditions,
e.g., the dusty circumgalactic medium around quasars or superluminous galaxies.
\end{abstract}

\begin{keywords}
cosmic rays --- plasmas --- methods: numerical --- MHD --- galaxies: active ---
ISM: structure
\end{keywords}

\section{Introduction}
\label{sec:intro}

The transport physics of cosmic rays (CRs) has been a subject of active
investigation since the 1960s \citep{kulsrud1969effect}. As charged
ultra-relativistic particles coupled to magnetic fields via Lorentz forces, CRs
are fundamentally governed by particle-wave-interactions with magnetic
fluctuations (e.g., Alfv{\'e}n waves). Generally speaking, when Alfv{\'e}n waves
are excited with wavelengths broadly similar to the CR gyro radii, CRs are
scattered toward isotropy in the wave frame from small-scale irregularities of
field lines and thus become ``confined'' (the net drift/streaming speed relative
to the plasma is suppressed). The relevant confining Alfv{\'e}n waves can be
excited by the CR streaming instability when the CR drift velocity $v_D$ exceeds
local Alfv{\'e}n velocity $v_\mathrm{A}\equiv |\bm{B}|/\sqrt{4\pi
\rho_\mathrm{gas}}$ \citep{kulsrud1969effect}, and/or by
extrinsic turbulence \citep{skilling1971cosmic,jokipii1966cosmic}. On the other
hand, CRs are less confined when Alfv{\'e}n waves are strongly damped, via
ion-neutral damping, nonlinear Landau damping \citep{lee1973damping} or through
the magnetohydrodynamic (MHD) turbulence cascade
\citep{yan2002scattering,farmer2004wave}. Therefore, studying the excitation and
damping mechanisms of Alfv{\'e}n waves around these gyro-resonant scales is
crucial to understanding of CR transport.

Recently, a number of numerical studies have attempted to model the effects of
CRs around energies of $\sim \mathrm{GeV}$ (which dominate the CR energy
density) on ISM and galactic scales. These studies suggest that CRs can have a
significant influence on galactic ``feedback processes'' regulating star and
galaxy formation (e.g.,
\citealt{pakmor2016galactic,ruszkowski2017global,farber2018impact,chan2019cosmic,hopkins2019but,buck2020effects,su2020cosmic})
and the phase structure and nature of the circumgalactic medium (CGM) (e.g.,
\citealt{salem2016role,butsky2018role,ji2020properties,ji2021virial}).
Meanwhile, classical models of Galactic CR transport which compare to Solar
System CR experiments
\citep{strong:2001.galprop,2016ApJ...824...16J,evoli:dragon2.cr.prop} have
suggested the potential for new breakthroughs in particle physics from detailed
modeling of ratios of e.g.\ secondary-to-primary ratios in CR populations.
However, in all of these studies a fluid or Fokker-Planck description of CR
transport is required, usually with some simple assumption that CRs have a
constant ``diffusion coefficient'' or ``streaming speed'' or effective
scattering rate. But this introduces significant uncertainties, as the
micro-physical behavior and even the qualitative physical origin of these
scattering rates (and therefore their dependence on plasma properties) remains
deeply uncertain. This uncertainty is vividly demonstrated in
\citet{hopkins2020testing}: comparing different CR transport scalings based on
local plasma properties according to different proposed scattering-rate models
in the literature, \citet{hopkins2020testing} showed that existing models could
(1) differ by factors larger than $\sim 10^{6}$ in their predicted effective
transport coefficients, and (2) even models with similar mean diffusion
coefficients could produce (as a  consequence of different detailed dependence
on plasma properties) qualitatively different CR transport outcomes and effects
on ISM and CGM properties. Therefore, it is particularly important to go beyond
simple fluid treatments of CRs by investigating explicit CR scattering physics
and dynamics on the scales of the CR gyro-radius $r_\mathrm{gyro,cr}$ ($\sim
\mathrm{AU}$ for $\mathrm{GeV}$ CRs in typical Solar-neighborhood ISM
conditions), where the interaction between CRs and magnetic fluctuations can be
fully resolved.

One potentially important piece of physics on these scales, which has been
almost entirely neglected in the historical CR literature, is the role of dust
grains. Recently, \citet{squire2020impact} noted a remarkable coincidence: under
a broad range of ISM conditions,  the gyro-radii of charged dust grains in the
ISM (with sizes $\sim 10^{-3}-1\,{\mathrm \mu m}$) and $\sim0.1-10$\,GeV CRs
overlap. While the grains have much lower charge-to-mass ratios, they also have
much lower velocities, giving nearly coincident gyro radii. As a result,
\citet{squire2020impact} proposed that dust grains can influence $\sim
\mathrm{GeV}$ CR transport on gyro-resonant ($\sim \mathrm{AU}$) scales in two
ways: (1) as inertial particles, dust can damp parallel Alfv{\'e}n waves excited
by the CR streaming instability and thus reduce CR confinement, and (2) dust can
be unstable to the so-called ``resonant drag instabilities'' (RDIs;
\citealt{squire.hopkins:RDI}), a recently-discovered, formally infinitely-large
family of instabilities  that appear in different forms, whenever dust grains
move through a background fluid with some non-vanishing difference in the force
acting on dust vs.\ fluid. The RDIs can excite small-scale parallel Alfv{\'e}n
waves in magnetized gas \citep{hopkins2018ubiquitous,seligman2019non}, which can
in turn scatter CRs and enhance their confinement. Which of these is the
dominant process, or whether still different forms of the RDIs can be excited,
depend on the local plasma conditions, with scenario(s) being more likely in
regions where the gas is relatively diffuse and external acceleration of dust
grains is relatively strong (from e.g.\ absorbed photon momentum around a bright
source). In any case, the co-existence of CRs and charged dust grains in the ISM
is self-evident. Even in the CGM far from galaxies, where
\citet{squire2020impact} suggest scenario (2) would be more likely, the
existence of significant dust populations is theoretically and observationally
plausible: for instance, in the cool ($\sim 10^4 - 10^5\,\mathrm{K}$) or
warm-hot ($10^5 - 10^6\,\mathrm{K}$) phases of the CGM where the gas number
density is low enough ($\lesssim 10^{-3}\,\mathrm{cm^{-2}}$), the hadronic and
Coulomb losses of CRs are small \citep{guo.oh:cosmic.rays}, and the temperature
and density are low enough that the dust sputtering time is long
\citep{tielens1994physics}. In fact, a significant amount of dust has already
been observed in the CGM \citep{menard2010measuring,peek2015dust}. And around
quasars or superluminous galaxies, dust grains could in principle easily be
accelerated by radiation pressure and become unstable to the specific RDIs which
would source strong CR scattering on $\sim \mathrm{AU}$ scales according to
\citet{squire2020impact}. Dust clumping can scatter and re-emit photons from
QSOs (e.g., \citealt{hennawi2010binary,martin2010size}), and thus might produce
observables in absorption lines of the CGM. Nevertheless, this scenario remains
largely unexplored; even for the pure dust RDIs without CRs, the only simulation
study thus far, in \citet{hopkins2020simulating}, has simulated the CGM cases at
a resolution of $\sim 10\,\mathrm{pc}$, far above $\mathrm{AU}$ scales on which
CR-dust coupling occurs.

Motivated  by the above, in this paper we present the very first numerical study
of CR-dust-MHD coupling, with resolved CR and dust gyro-radii. In particular,
here we focus on the proposed CR confinement scenario in which CRs are scattered
by magnetic field irregularities caused by the specific forms the RDI described
in \citet{squire2020impact}. The paper is organized as follows. In
\S\ref{sec:methods}, we describe the numerical methods and setup of our
simulations. \S\ref{sec:results} presents and analyzes the simulation results.
We finally summarize our findings in \S\ref{sec:conclusions}.

\section{Methods \& Simulation Setup}
\label{sec:methods}

\subsection{Numerical Methods}

The numerical methods adopted here consist of three main components, each of
which has been well-studied separately: the MHD solver in the code {\small
GIZMO} \citep{hopkins:gizmo}, dust grains evolved with a standard super-particle
method \citep{hopkins2018ubiquitous,seligman2019non}, and CR particles evolved
with a hybrid MHD particle-in-cell (MHD-PIC) method
\citep{bai2015magnetohydrodynamic,bai2019magnetohydrodynamic}. Detailed
implementations are described in the above-cited references, and we only briefly
review them here. 

The background plasma/fluid/gas and magnetic fields follow the equations of
ideal MHD,  
solved in {\small GIZMO} with the well-tested constrained-gradient meshless
finite-volume Lagrangian Godunov method
\citep{hopkins:mhd.gizmo,hopkins:cg.mhd.gizmo}. This has been shown to
accurately reproduce a variety of detailed MHD phenomena including
amplification, shocks, detailed structure of the magnetorotational and
magnetothermal instabilities, and more
\citep{hopkins:gizmo.diffusion,deng:2019.mri.turb.sims.gizmo.methods,grudic:starforge.methods}.
To this we add the ``back-reaction'' or feedback force of CRs and dust on gas,
detailed below. Individual grains and CRs are integrated as PIC-like
``super-particles'' as is standard in the literature for both
\citep{carballido:2008.grain.streaming.instab.sims,johansen:2009.particle.clumping.metallicity.dependence,bai:2010.grain.streaming.vs.diskparams,pan:2011.grain.clustering.midstokes.sims,2018MNRAS.478.2851M,bai2015magnetohydrodynamic,bai2019magnetohydrodynamic,holcolmb.spitkovsky:saturation.gri.sims,2019MNRAS.tmp.2249V},
sampling the distribution function statistically by taking each super-particle
to represent an ensemble of identical micro-particles (individual grains or
CRs). Dust grains obey the equation of motion 
\begin{align}
   \frac{d\bm{v}_\mathrm{dust}}{dt} = \bm{a}_\mathrm{ext,dust} + \bm{a}_\mathrm{gas,dust}, 
\end{align}
where $\bm{v}_\mathrm{dust}$ is the grain velocity, $\bm{a}_\mathrm{ext,dust}$
and $\bm{a}_\mathrm{gas,dust}$ are grain accelerations from external forces
(e.g.\ gravity, radiation) and background gas and magnetic fields respectively.
The latter includes both drag  and Lorentz forces:
\begin{align}
  \bm{a}_\mathrm{gas,dust} = - \frac{\bm{v}_\mathrm{dust} - \bm{v}_\mathrm{gas}}{t_{s,\mathrm{dust}}} - \frac{\left(\bm{v}_\mathrm{dust} - \bm{v}_\mathrm{gas}\right) \times \hat{\bm{B}}}{t_{L,\mathrm{dust}}},
\end{align}
where $\hat{\bm{B}}=\bm{B}/|\bm{B}|$ is the unit vector of the magnetic field
$\bm{B}$, $t_{s,\mathrm{dust}}$ the dust drag drag or stopping time and
$t_{L,\mathrm{dust}}$ the dust Larmor or gyro time. Since we are interested in
regimes with super-sonic drift and microscopic dust grains (the Epstein drag
limit), the drag and gyro timescales are given by:
\begin{align}
  t_{s,\mathrm{dust}} &\equiv \sqrt{\frac{\pi \gamma}{8}} \frac{\rho_{\mathrm{dust}}^{i}}{\rho_\mathrm{gas}} \frac{\epsilon_d}{c_s} \left(1 + \frac{9 \pi \gamma}{128} \frac{\left|\bm{v}_\mathrm{dust} - \bm{v}_\mathrm{gas}\right|^2}{c_s^2}\right)^{-1/2} \\
  t_{L,\mathrm{dust}} &\equiv \frac{m_\mathrm{dust} c}{q_\mathrm{dust} \bm{B}} = \frac{4\pi \rho_\mathrm{dust}^i \epsilon_\mathrm{dust}^3 c}{3 e \left|Z_\mathrm{dust} \bm{B}\right|},
\end{align}
where $\gamma$ is the usual adiabatic index of the gas,  and
$\rho_{\mathrm{dust}}^{i}$, $\epsilon_\mathrm{dust}$, $m_\mathrm{dust}$ and
$q_\mathrm{dust}\equiv Z_\mathrm{dust} e$ are the internal density, radius, mass
and charge of dust grains respectively.

Similarly, the equation of motion for CRs is:
\begin{equation}
\label{eqn:cr.accel} \left( \frac{c}{\tilde{c}} \right)\frac{d\bm{u}_\mathrm{cr}}{dt} = \bm{a}_\mathrm{ext,cr} +  \bm{a}_\mathrm{gas,cr}  =  \bm{a}_\mathrm{ext,cr}  + \frac{\left(\bm{v}_\mathrm{cr} - \bm{v}_\mathrm{gas}\right)\times \hat{\bm{B}}}{t_{L,0,\mathrm{cr}}},
\end{equation}
where $c$ is the speed of light (and see below for $\tilde{c}$), and
$\bm{v}_\mathrm{cr}$ is the velocity of the CR particles, which is related to
the CR four-velocity $\bm{u}_\mathrm{cr} \equiv \gamma_L \bm{v}_\mathrm{cr}$ via
the usual Lorentz factor $\gamma_L \equiv ( 1 -  |\bm{v}_\mathrm{cr}|^2 /
c^{2})^{-1/2}$. The $t_{L,0,\mathrm{cr}} \equiv 1/\Omega_\mathrm{cr,0}$ is the
usual non-relativistic CR gyro time $t_{L,0,\mathrm{cr}} \equiv (m_\mathrm{cr}
c)/(q_\mathrm{cr} \bm{B})$, 
where $m_\mathrm{cr}$ and $q_\mathrm{cr}$ are the CR mass and charge.

In the ideal MHD approximation, the ``feedback'' force from dust grains and CRs
appears simply in the gas momentum equation as an equal-and-opposite force as
from gas onto dust+CRs:
\begin{align}
  \rho_\mathrm{gas} &\left(\frac{\partial}{\partial t}+ \bm{v}_\mathrm{gas} \cdot \nabla \right) \bm{v}_\mathrm{gas} 
  = -\nabla P - \frac{\bm{B}\times (\nabla \times \bm{B})}{4\pi} \notag \\
  & \qquad - \int d^3 \bm{v}_\mathrm{dust} f_\mathrm{dust} (\bm{x}, \bm{v}_\mathrm{dust})\,\bm{a}_\mathrm{gas,dust}(\bm{v}_\mathrm{dust},...) \notag \\
  & \qquad \qquad \qquad \qquad - \int d^3 \bm{v}_\mathrm{cr} f_\mathrm{cr} (\bm{x}, \bm{v}_\mathrm{cr})\,\bm{a}_\mathrm{gas,cr}(\bm{v}_\mathrm{cr},...),
  \label{eq:momentum}
\end{align}
where $f_\mathrm{dust} (\bm{x}, \bm{v}_\mathrm{dust})$ and $f_\mathrm{cr}
(\bm{x}, \bm{v}_\mathrm{cr})$ are the phase-space density distributions of dust
and CRs respectively. Non-Lorentz forces are integrated with a semi-implicit
scheme and Lorentz forces using a Boris integrator, with the back-reaction terms
implemented in a manner that ensures manifest machine-accurate total momentum
conservation. These methods have been detailed and extensively tested in
\citet{hopkins.2016:dust.gas.molecular.cloud.dynamics.sims,lee:dynamics.charged.dust.gmcs,moseley2019non,seligman2019non,hopkins2020simulating}.

Note that up to the detailed form of the gyro acceleration equation, the
expressions for dust and gas evolution are functionally identical -- indeed we
can numerically think of the dust as a second, ``heavy'' (low charge-to-mass,
non-relativistic) CR species which also experiences drag, or think of the CRs as
``relativistic, drag-free grains.''

Finally, to avoid the Courant condition for the CR speed $c$ leading to
computationally impractical timesteps in our simulation, we adopt the  reduced
speed-of-light (RSOL) approximation by defining the RSOL $\tilde{c}$ in
Eq.~\eqref{eqn:cr.accel} $\tilde{c} < c$ (but still keeping $\tilde{c}$ much
larger than other velocities in our simulation). As shown in
\citet{ji2021accurately}, the particularly form of the RSOL implemented here,
which simply modifies the CR acceleration by the power of $\tilde{c}/c$ and
defines the CR advection speed over the (non-relativistic) grid to be
$\bm{v}_\mathrm{cr}^\mathrm{advect} \equiv \partial \bm{x}_\mathrm{cr} /
\partial t = (\tilde{c}/c)\,\bm{v}_\mathrm{cr}$, ensures that all steady-state
properties of the CR distribution function (number, momentum, energy density,
pitch-angle distribution) are mathematically invariant to the choice of
$\tilde{c}$. With this new implementation, we are able to simultaneously match
the CR energy density, mass density, \emph{and} momentum density of any desired
initial configuration, and account for the ``correct'' CR gyro-radius and
back-reaction forces which are also independent of the choice of $\tilde{c}$.
This also means that throughout this paper when we refer to e.g.\ CR
gyro-frequencies and other standard quantities, they have their usual meaning
and values (i.e.\ unless otherwise specified, the RSOL $\tilde{c}$ does not
enter our expressions here).

\subsection{Simulation Parameters \&\ Motivation}
\label{sec:parameters}

In what follows, we consider a system where dust grains are accelerated by some
large $\bm{a}_\mathrm{ext,dust}$, for example from radiation pressure from e.g.\
a quasar or luminous starburst in their host galaxy, in low-density magnetized
gas representative of e.g.\ the CGM or ionized ISM bubbles. In these cases the
fastest growing of the dust RDIs are generally the ``aligned cosmic-ray like''
(so named because the eigenmode structure broadly resembles the
\citealt{bell.2004.cosmic.rays} instability) or ``dust gyro-resonant'' RDIs,
which lead to rapid growth of parallel Alfv{\'e}n waves which can scatter CRs.
For the cases of interest, the external forces on the CRs
$\bm{a}_\mathrm{ext,cr}$ are negligible.

Because this is a first study, to simplify the dynamics as much as possible and
have a well-defined, resolved CR and dust gyro-radius, we consider just a single
species of dust and single species of CRs. In reality of course there will exist
a broad spectrum of dust sizes, with different charge and mass, e.g., and
likewise spectrum of CR energies; but we will focus on parameters representative
of the grains that contain most of the dust mass and dominate the ``feedback''
force, as well as the CRs which dominate the CR energy density/pressure. We
adopt a 3D box with a side-length of $L_\mathrm{box}$ and periodic boundary
conditions, filled with initially homogeneous gas, dust, and CRs, with a uniform
magnetic field whose initial direction defines the $z$ axis ($\bm{B}_{0} =
|\bm{B}_{0}|\,\hat{z}$) and uniform velocity fields (i.e.\ the initial CR and
dust distribution functions are taken to be $\delta$-functions, though we
discuss relaxing this below), and impose an isothermal $(\gamma=1$)
equation-of-state on the gas (motivated by typical cooling physics in the
ISM/CGM). We adopt the same particle/cell numbers for gas, dust and CRs, i.e.,
$N_\mathrm{gas} = N_\mathrm{dust} = N_\mathrm{cr} = N_\mathrm{1D}^3$, where
$N_\mathrm{1D}$ is the 1D resolution along each sides of the simulation box,
$N_\mathrm{1D}=64$ in our fiducial simulations (so the box contains $3\times
64^{3}$ elements), and we also performed additional high-resolution simulations
with $N_\mathrm{1D}=128$ for convergence tests. We initialize the dust drift
velocity with its homogeneous equilibrium solution (see
\citealt{hopkins2020simulating}), the gas velocity $\bm{v}_\mathrm{gas} = 0$,
and the CR velocity $\bm{v}_\mathrm{cr} = v_\mathrm{cr}\,\hat{z}$, defined
below. 

Even with these simplifications, writing the simulation parameters in units of
the initial gas density $\rho_\mathrm{g}^0$, sound speed $c_s^0$ and box size
$L_\mathrm{box}$, our simulations require we specify ten dimensionless numbers:
(1) The plasma beta $\beta \equiv (c_s / v_\mathrm{A})^2$, where
$v_\mathrm{A}\equiv \bm{B}/\sqrt{4\pi \rho_\mathrm{gas}}$ is the Alfv{\'e}n
velocity;
(2) The box-averaged dust-to-gas mass ratio $\mu_\mathrm{dust} \equiv
\rho_\mathrm{dust} / \rho_\mathrm{gas}$;
(3) The dust external acceleration $\bar{a}_\mathrm{dust}\equiv
|\bm{a}_\mathrm{ext,dust}| L_\mathrm{box} / (c_s^0)^2$; 
(4) The dust ``size parameter'' $\bar{\epsilon}_\mathrm{dust} \equiv
\rho_\mathrm{dust}^i \epsilon_\mathrm{dust} / \rho_\mathrm{g}^0 L_\mathrm{box}$,
which determines the grain drag forces;
(5) The dust ``charge parameter'' $\bar{\phi}_\mathrm{dust} \equiv 3
Z_\mathrm{dust} e / (4\pi c \epsilon_\mathrm{dust}^2
(\rho_\mathrm{gas}^0)^{1/2})$, which determines the grain charge-to-mass ratio
and Lorentz forces; 
(6) The angle $\mathrm{cos}(\theta_\mathrm{dust}) \equiv \left|\hat{\bm{B}}_{0}
\cdot \hat{\bm{a}}_\mathrm{ext,dust}\right|$ between initial $B$-field direction
and external dust acceleration; 
(7) The CR-to-gas mass ratio $\mu_\mathrm{cr} \equiv \rho_\mathrm{cr} /
\rho_\mathrm{gas}$ (or equivalently the CR number ratio
$n_\mathrm{cr}/n_\mathrm{gas}$, since we are interested in CR protons);
(8) The CR ``charge parameter'' $\bar{\phi}_\mathrm{cr}\equiv q_\mathrm{cr} /(c
m_\mathrm{cr}) L_\mathrm{box} (\rho_\mathrm{gas}^0)^{1/2}$, which encodes the CR
charge-to-mass ratio;
(9) The angle $\mathrm{cos}(\theta_\mathrm{cr}) \equiv \left|\hat{\bm{B}}_{0}
\cdot \hat{\bm{v}}_\mathrm{cr}\right|$ between the initial magnetic fields and
CR velocity.
(10) The initial CR Lorentz factor $\gamma_L$ (or equivalently initial CR
momentum $p_\mathrm{cr}$). 

This forms an enormous parameter space, which is impossible to explore
concisely. We therefore focus on one particular parameter set in this first
study, motivated heuristically by the scenario proposed in
\citet{squire2020impact} and discussed above. Consider typical
interstellar/intergalactic silicate or carbonaceous dust ($\rho_\mathrm{dust}^i
\sim 1\,\mathrm{g\,cm^{-3}}$, with $\epsilon_\mathrm{dust} \sim 0.1\,\mathrm{\mu
m}$ grains containing most of the mass, a standard ISM-like dust-to-gas ratio,
and obeying the standard collisional+photoelectric charge law from e.g.\
\citealt{draine:1987.grain.charging,tielens:2005.book}) and typical $\sim1\,$GeV
kinetic energy CRs which dominate the CR energy density ($\gamma_{L}\sim2$
protons). These proposes in a medium with parameters typical of the warm
(volume-filling) CGM outside of a galaxy (gas temperature $T \sim
10^5\,\mathrm{K}$, density $n\sim 10^{-2}\,\mathrm{cm^{-3}}$, $|\bm{B}| \sim
0.1\,\mathrm{\mu G}$, and CR-to-thermal pressure $P_\mathrm{cr}/P_\mathrm{therm}
\sim$\,a few; \citealt{ji2020properties}). We wish to resolve some number of CR
gyro radii in our box, so set $L_\mathrm{box} \sim
10\,r_{L,\mathrm{cr}}$.\footnote{More precisely, we set
$L_\mathrm{box}=\sqrt{N_\mathrm{1D}}\,r_{L,\,\mathrm{cr}}$ such that each Larmor
wavelength is resolved with $\sim \sqrt{N_\mathrm{1D}}$ elements and the box
contains $\sqrt{N_\mathrm{1D}}$ Larmor wavelengths, which provides an optimal
compromise for our purposes. For boxes with different resolution, we rescale the
numerical parameters to correspond to fixed {\em physical} quantities (e.g.\
fixed gyro radii) while rescaling the box size to follow this relation. Here and
throughout, we define the gyro-radius specifically as the gyro radius for an
equivalent circular orbit, $r_{L,\mathrm{cr}} =
|\bm{v}_\mathrm{cr}|\,t_{L,\mathrm{cr}}$.} This gives the numerical
parameters\footnote{For our low-resolution $N_\mathrm{1D}=64$ boxes, the
parameters which depend explicitly on $L_\mathrm{box}$ rescale to
$\bar{\epsilon}_\mathrm{dust} \approx 1.5\times 10^5$, $\bar{\phi}_\mathrm{cr}
\approx 9\times10^4$, $\bar{a}_\mathrm{dust} \approx 8.5\times 10^{-4}$.} $\beta
\approx 2\times 10^3$, $\mu_\mathrm{dust} \approx 0.01$,
$\bar{\epsilon}_\mathrm{dust} \approx 10^5$, $\bar{\phi}_\mathrm{dust}
\approx1.6\times 10^6$, $\mu_\mathrm{cr} \approx 3.6\times10^{-7}$,
$\bar{\phi}_\mathrm{cr} \approx 1.3\times10^5$,  $\gamma_L\approx
2$.\footnote{We also set reduced speed of light to $\tilde{c} = 0.05\,c \approx
230 c_s^0$, ensuring it is larger than other speeds in the problem, but
explicitly examine convergence with respect to this choice below.} What remains
is the external dust acceleration $\bar{a}_\mathrm{dust}$, which is the ultimate
source of energy for the RDIs: a larger value of this parameter corresponds to
more rapid RDI growth rates. \citet{squire2020impact} considered super-sonic
dust drift velocities ($v_\mathrm{drift} \equiv
|\bm{v}_\mathrm{dust}-\bm{v}_\mathrm{gas}|$), such that certain RDIs were
excited, so we choose $\bar{a}_\mathrm{dust} = 1.2\times 10^{-3}$ such that the
initial equilibrium dust drift velocity is $v_\mathrm{drift} \approx 10\,c_s^0$
(safely super-sonic).\footnote{As there is no particular reason to expect
alignment between $\bm{B}$ and $\bm{a}_\mathrm{ext,dust}$, we set
$\theta_\mathrm{dust} = 45^\circ$, but this is largely a nuisance parameter.} As
discussed in \citet{hopkins2018ubiquitous,hopkins2020simulating} and
\citet{squire2020impact}, while this is intentionally a somewhat extreme choice,
it is plausible given the observed radiative fluxes in CGM environments around
super-luminous sources such as quasars and starburst galaxies, e.g., at
$\sim10-100\,$kpc from a source with luminosity $\sim
10^{12}-10^{13}\,L_{\odot}$.

It is easy to verify that under these conditions, Lorentz forces on dust grains
strongly dominate over drag forces (by a factor of $\sim 10^{4}$),  as desired
for our problem of interest, and that both CRs and dust have very similar
gyro-radii, $\sim 0.6\,\mathrm{AU}$ in the physical units above.

\subsection{On the Initial CR Pitch Angle Distribution}

Note that our initial condition for the CRs corresponds to all CRs having pitch
angle $=0$ ($\mu=1$), i.e.\ free-streaming directly down magnetic field lines at
drift velocity $v_D^0 \sim c$ with the maximum-possible anisotropy in the CR
distribution function. This is almost certainly unrealistic, but our purpose is
to study how CRs would be scattered away from this anisotropy by dust, hence the
choice.\footnote{In contrast, because of drag, the only homogeneous stable
equilibrium solution for the dust grains is uniform streaming at the equilibrium
drift velocity, as we initialize.} However one consequence of this choice is
that for the parameters above (with  $(e_\mathrm{cr} v_D^0) / (e_B c) > 1$,
where $e_\mathrm{cr}$ and $e_B$ are the CR and magnetic energy densities), the
non-resonant Bell instability would grow much faster than the CR resonant
instability \citep{bell1978acceleration,haggerty2019hybrid} and much faster than
the RDIs (by a factor $\sim (\mu_\mathrm{cr}/\mu_\mathrm{dust})^{1/2} (v_D^0 /
v_\mathrm{dust}^0) (t_{L,\mathrm{cr}}^0/t_{L,\mathrm{dust}}^0)^{-1} \gg 1$). And
indeed we have  verified  this directly in test simulations, which also allow us
to confirm that, in code, both the non-resonant and resonant CR instabilities
grow initially at their expected linear growth rates. 

A more realistic initial condition would feature a close-to-isotropic CR
distribution function, which would strongly suppress these instabilities
relative to the RDIs, and  we also consider this case below. However for a first
simulation, in order to see how CRs would be scattered from an arbitrarily
anisotropic distribution with $\mu=1$ (which is particularly useful for
understanding the scattering physics themselves), we consider the ``ultra-low CR
density'' or ``test-particle CR'' limit. This amounts to to ignoring the CR
feedback on the gas (equivalently, taking the CR number or energy density to be
negligibly small), i.e., dropping the last term on the right-hand side in Eq.
\eqref{eq:momentum}. Thus, CR scattering via self-induced instabilities cannot
occur and  we can cleanly isolate the effects of the RDIs. We study this as a
test problem in \S\ref{sec:spatial} -- \S\ref{sec:without_feedback}. After the
system reaches a saturation state and CRs become more isotropized with $v_D \ll
\tilde{c}$ (a more realistic ``initial condition'' for the CRs), we turn on CR
feedback and investigate its consequences in \S\ref{sec:with_feedback}.

\section{Results}
\label{sec:results}

\subsection{Time Evolution \&\  Saturated  States}
\label{sec:spatial}

\begin{figure*}
  \begin{centering}
    \includegraphics[width=0.983\textwidth,left]{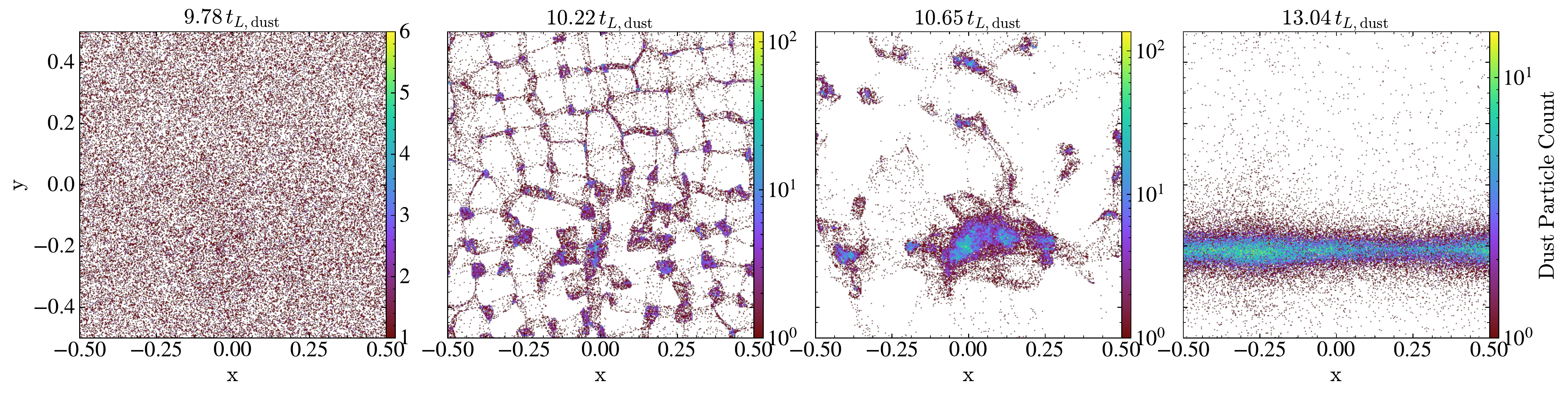}
    \includegraphics[width=\textwidth,left]{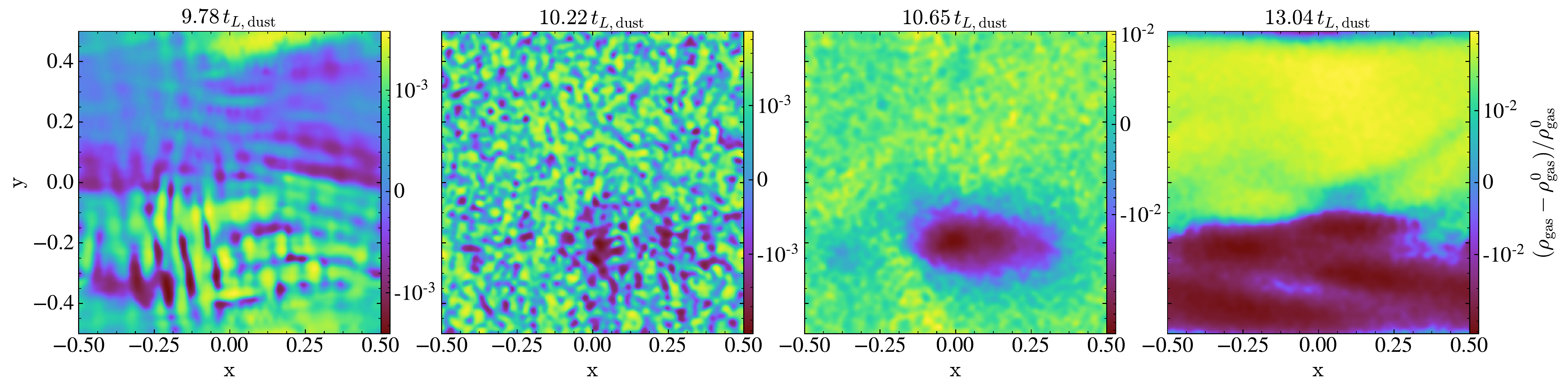}
    \includegraphics[width=\textwidth,left]{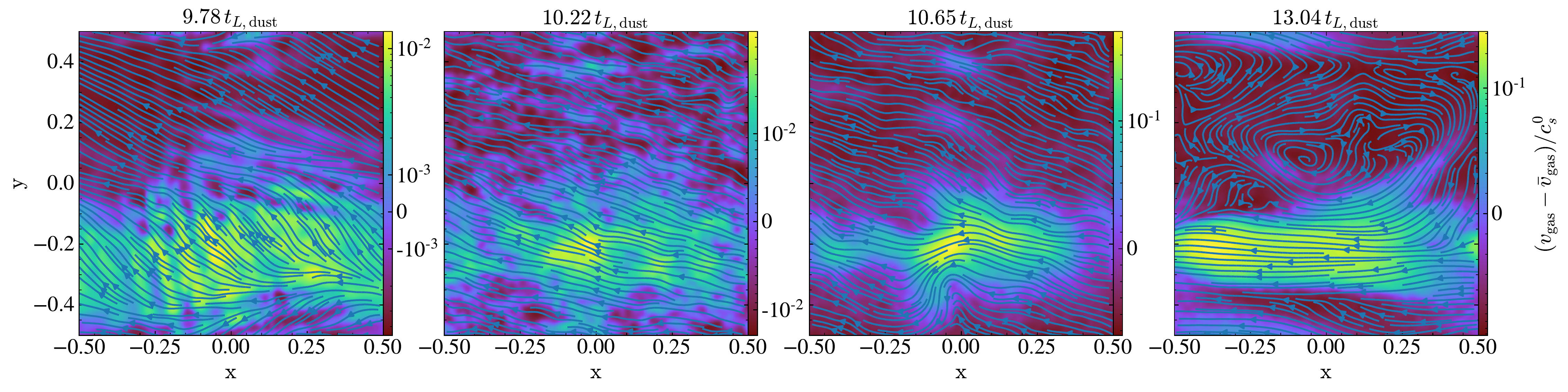}
    \includegraphics[width=\textwidth,left]{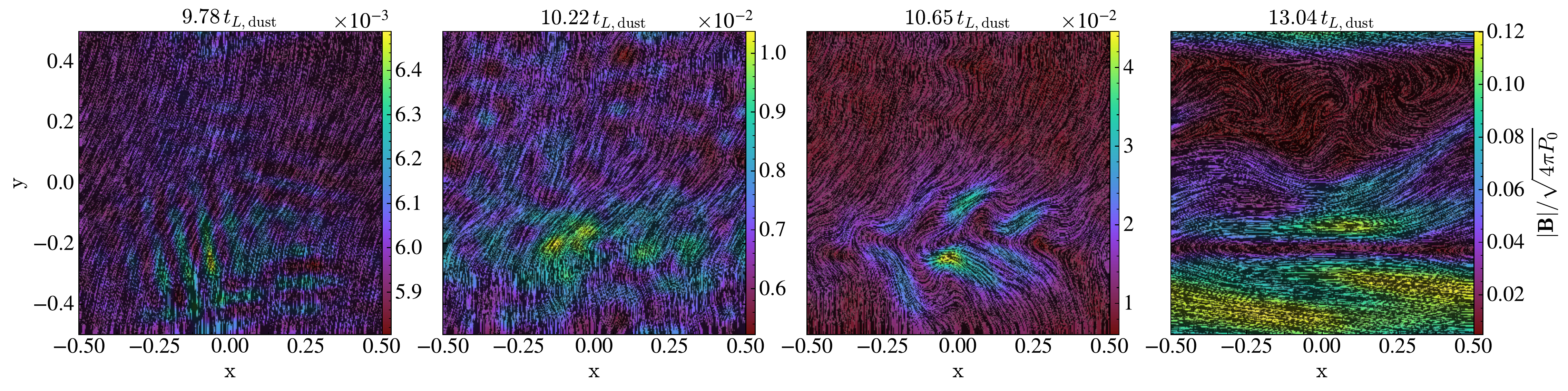}
    \includegraphics[width=0.983\textwidth,left]{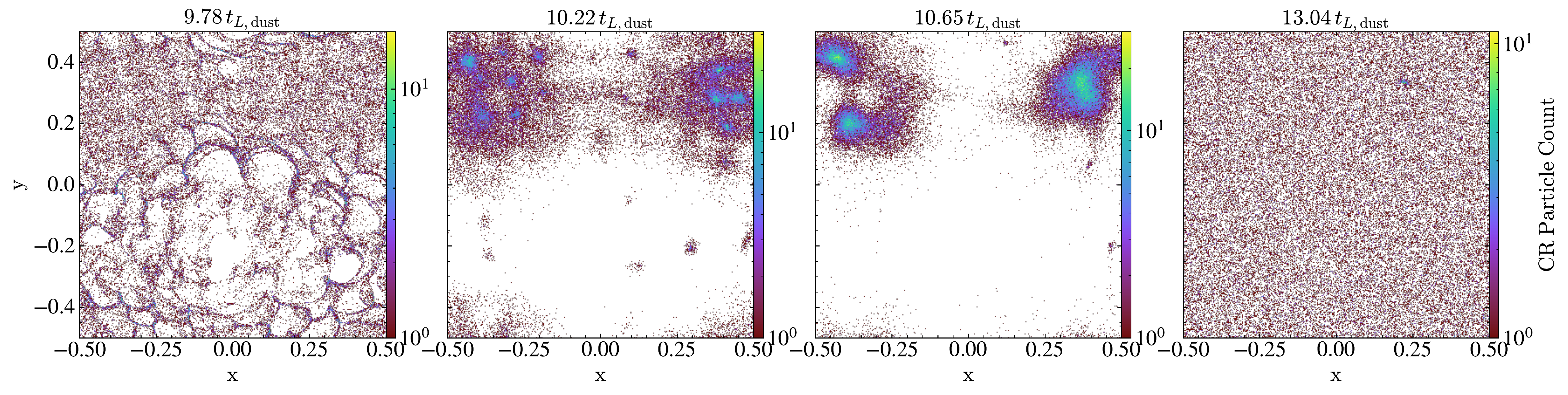}
  \end{centering}
  \vspace{-0.5cm}
  \caption{Plots of (from top to bottom) dust grain projections, gas density
  fluctuations, gas velocity perturbations superposed with velocity streamlines,
  magnetic field strengths superposed with field lines, and CR particle
  projections, at (from left to right) $t=9.78\,t_{L,\mathrm{dust}}$, $10.22
  \,t_{L,\mathrm{dust}}$, $10.65 \,t_{L,\mathrm{dust}}$ and $13.04\,
  t_{L,\mathrm{dust}}$, viewed along the $z$-axis (parallel with the direction
  of initial magnetic fields). Dust grains are unstable to the RDI and modes
  grow and merge until they saturate in a large box-scale sheet mode, which
  significantly distorts and amplifies magnetic fields. CRs strongly respond to
  and are scattered by dust-induced magnetic field irregularities.
  \label{fig:slice}\vspace{-0.2cm}}
\end{figure*}

\begin{figure*}
  \begin{centering}
    \includegraphics[width=0.983\textwidth,left]{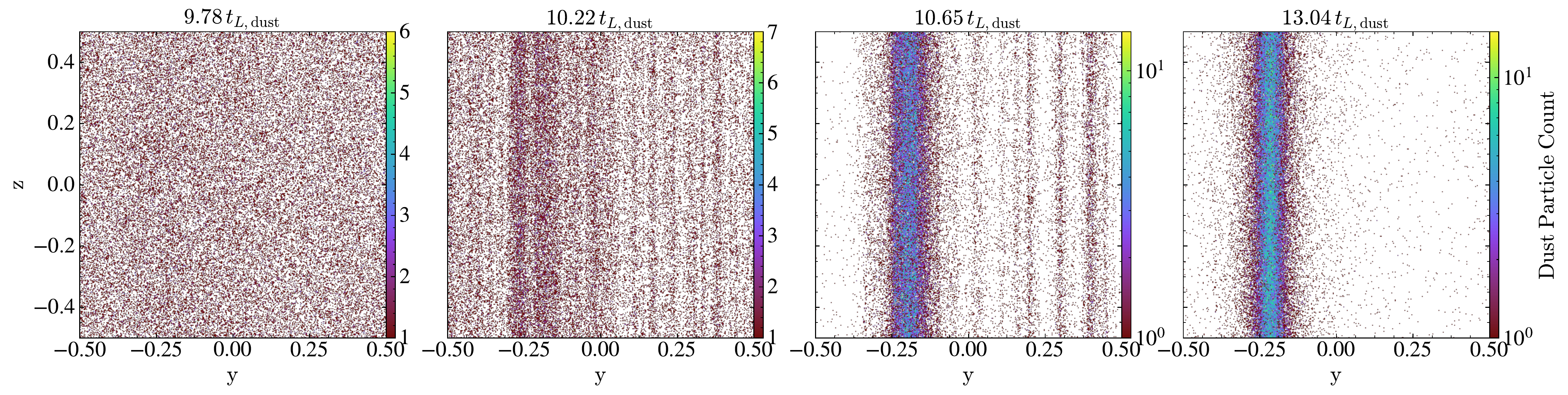}
    \includegraphics[width=\textwidth,left]{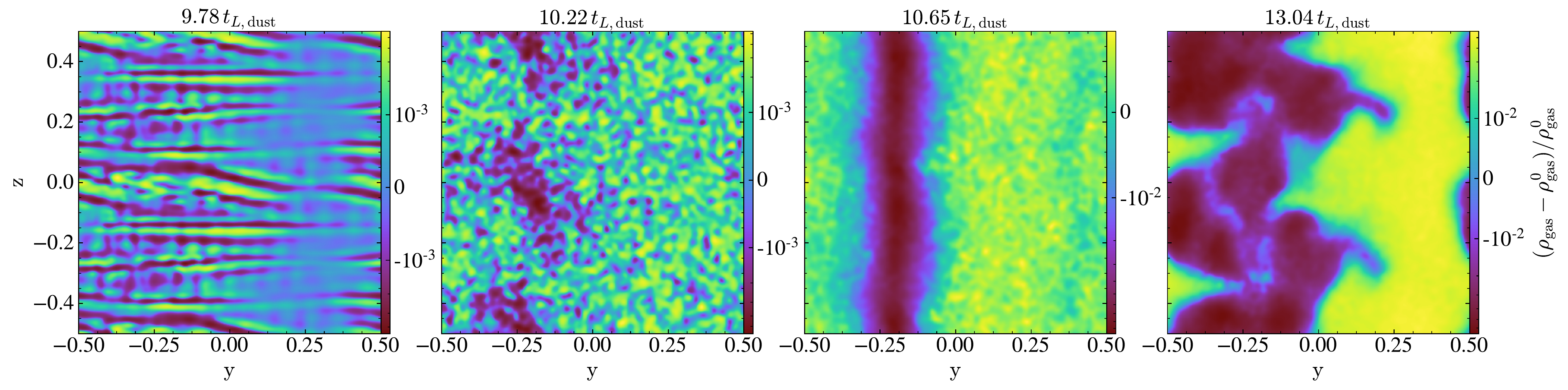}
    \includegraphics[width=\textwidth,left]{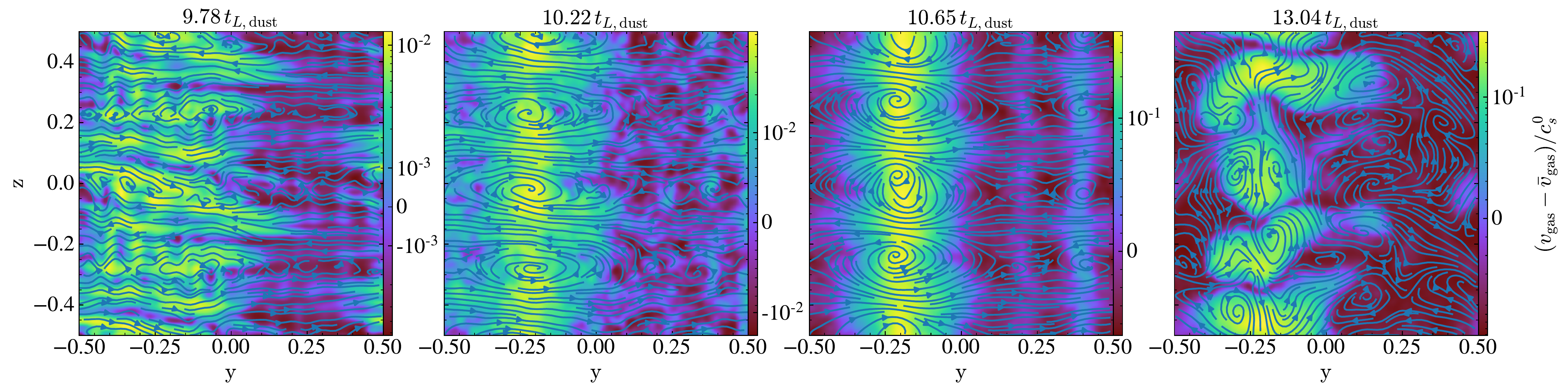}
    \includegraphics[width=\textwidth,left]{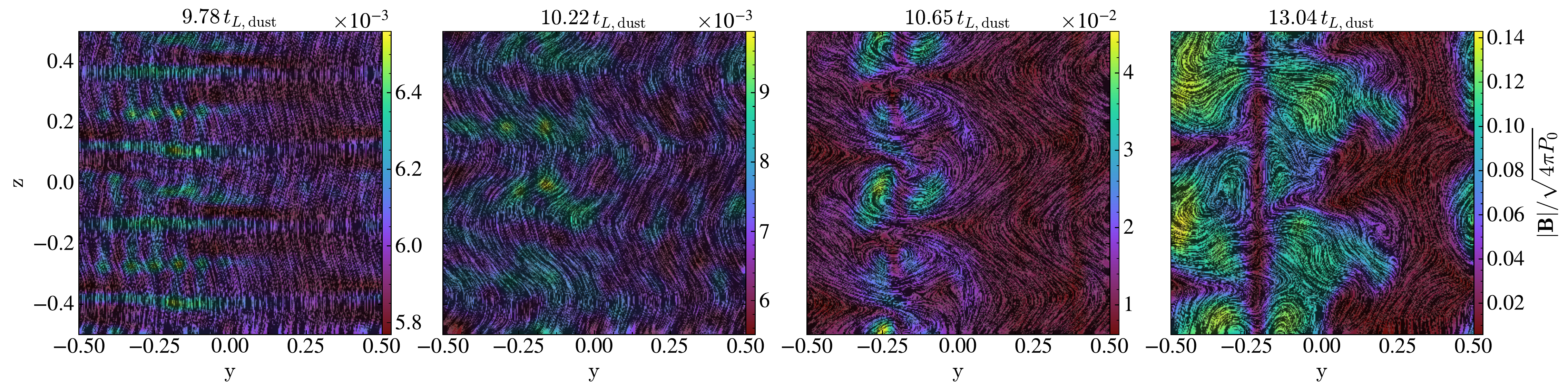}
    \includegraphics[width=0.983\textwidth,left]{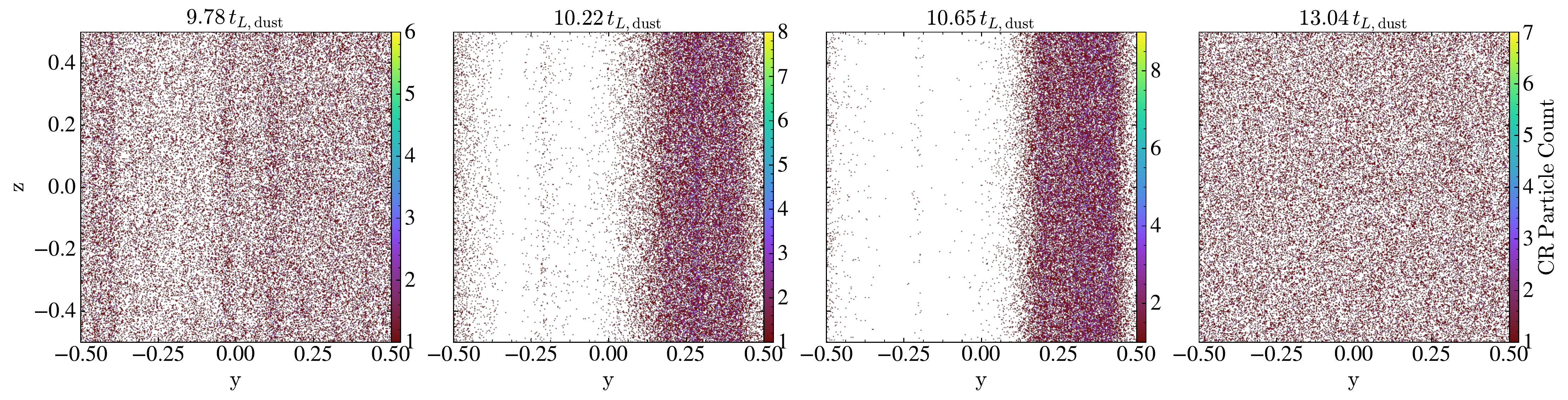}
  \end{centering}
  \vspace{-0.5cm}
  \caption{Plots as Fig. \ref{fig:slice}, but viewed along the $x$-axis
  (perpendicular to the direction of initial magnetic fields).
  \label{fig:slice_y}\vspace{-0.2cm}}
\end{figure*}

\begin{figure*}
  \begin{centering}
    \includegraphics[width=\textwidth]{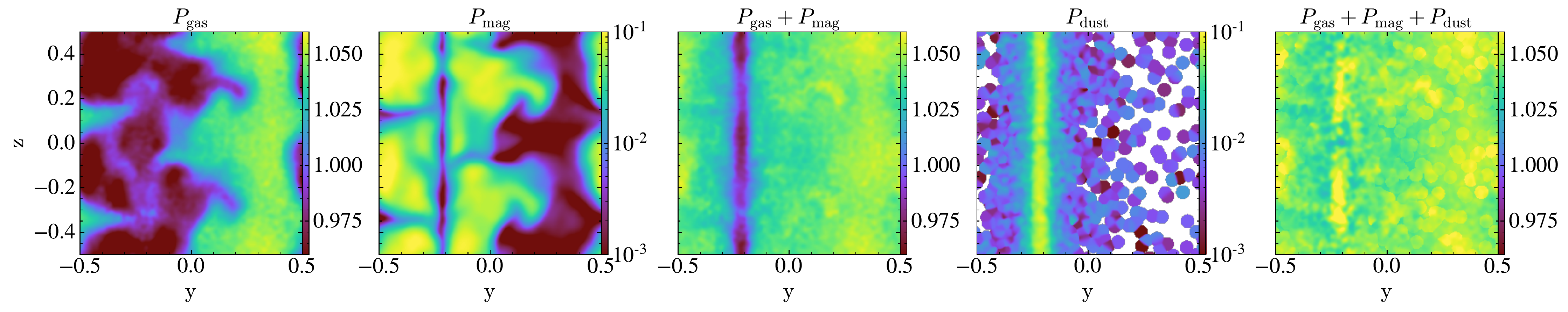}
  \end{centering}
  \vspace{-0.5cm}
  \caption{Slice plots of (from left to right) gas thermal pressure
  $P_\mathrm{gas}$, magnetic pressure $P_\mathrm{mag}$, the sum of gas thermal
  pressure and magnetic pressure $P_\mathrm{gas} + P_\mathrm{mag}$, dust ram
  pressure $P_\mathrm{dust}$ estimated as dust momentum flux across the surface
  of the dust filament $\rho_\mathrm{dust} v_{z,\mathrm{dust}}\,
  v_{xy,\mathrm{dust}}$, and the sum of all pressure terms $P_\mathrm{gas} +
  P_\mathrm{mag}+P_\mathrm{dust}$ in code units of $\rho_\mathrm{gas}^0
  (c_s^0)^2$, at the saturation stage when $t = 13.04\,t_{L,\mathrm{dust}}$. The
  sum of pressure terms is nearly a constant spatially, indicating the system is
  roughly in pressure balance. \label{fig:slice_balance}\vspace{-0.2cm}}
\end{figure*}

\begin{figure*}
  \begin{centering}
    \includegraphics[width=\textwidth]{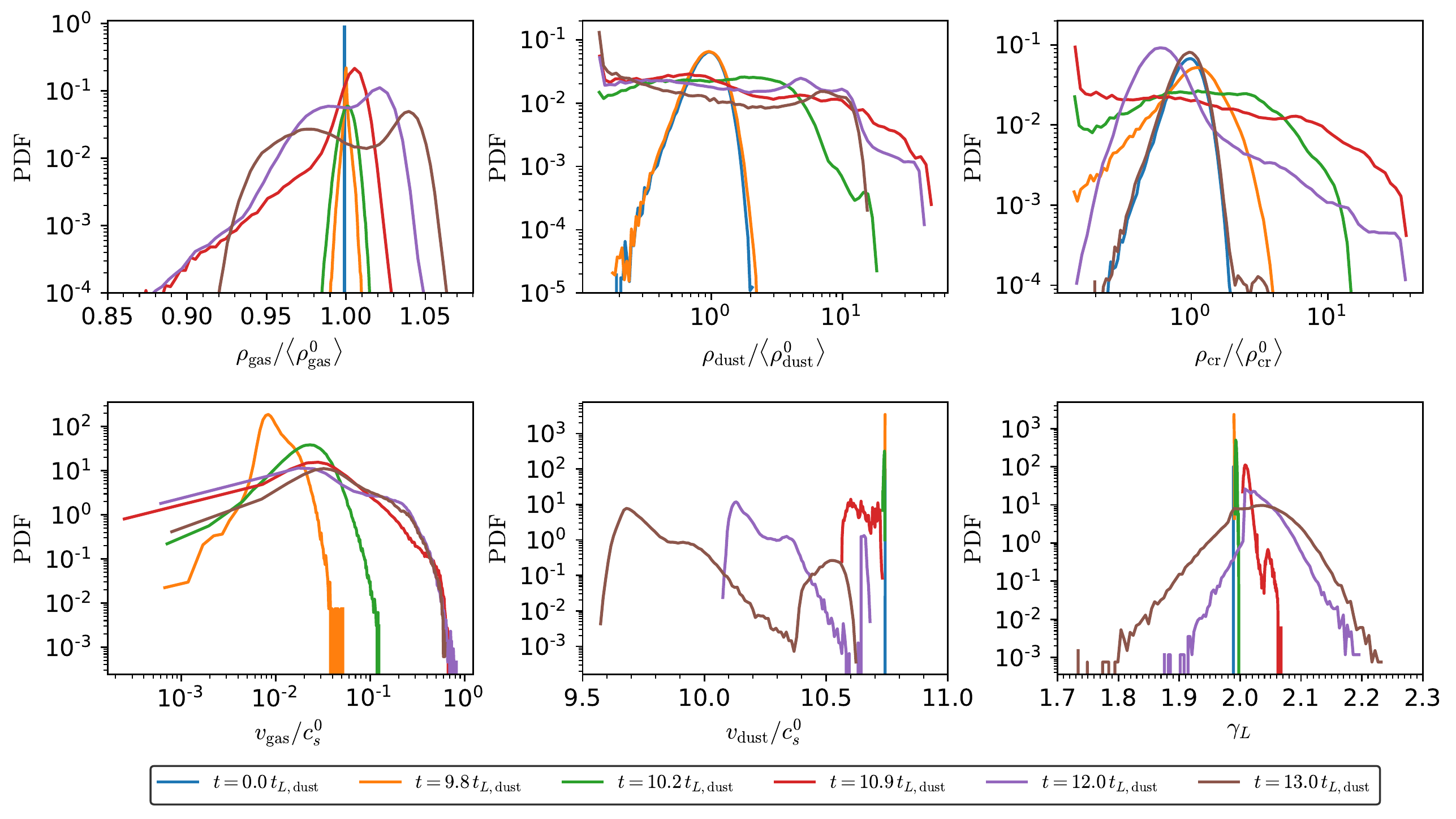}
  \end{centering}
  \vspace{-0.5cm}
  \caption{The density (top) and velocity / Lorentz factor (bottom) PDFs of gas
  (left), dust (middle) and CRs (right). Most PDFs feature a strong asymmetry.
  The density PDFs of dust and CRs span up to two orders of magnitude, while gas
  is only weakly compressible. Gas and dust grains are accelerated and
  decelerated respectively due to their momentum exchange, and the PDFs of CR
  Lorentz factor slightly broaden with time. \label{fig:PDF}\vspace{-0.2cm}}
\end{figure*}

Fig. \ref{fig:slice} and \ref{fig:slice_y} shows plots of the dust grain
projections, gas density fluctuations, gas velocity perturbations superposed
with velocity streamlines, magnetic field strength superposed with field lines,
and cosmic ray particle projections, viewed along the $z$-axis (the direction of
the initial magnetic fields) and $x$-axis respectively. We choose the snapshots
at $t=9.78 \,t_{L,\mathrm{dust}}$, $10.22 \,t_{L,\mathrm{dust}}$,
$10.65\,t_{L,\mathrm{dust}}$ and $13.04\, t_{L,\mathrm{dust}}$, which are
representative samples spanning between the end of the linear stage, the
nonlinear stage and the saturation stage. Dust grains are unstable to the RDI
and non-linearly evolve into highly-concentrated columns along the $z$-axis,
which merge and form into fewer but thicker columns or sheets with time. At the
saturation stage, all dust grains form into one single column in our simulation
box. This is expected since all wavelengths are unstable to RDIs, and the RDI
growth rates decrease with increasing wavelengths. Therefore, we see the merging
of structures until the box-scale mode saturates. Compared to collisionless dust
grains which are strongly clumped, gas is only weakly compressible, with density
and velocity fluctuations at levels of $\sim 1\%$ and $10\%$ respectively. As
shown in Fig. \ref{fig:slice} and \ref{fig:slice_y}, since field lines are
strongly stretched by dust grains, magnetic fields are amplified by up to one
order of magnitude from their initial values, and the regions of field
amplification tightly trace the location of clustered dust. At the saturation
stage, field lines become significantly distorted and wrap around columns of
dust grains, with the magnetic tension force $\sim \bm{B}\times (\nabla \times
\delta \bm{B}) / 4\pi$ balancing the driving force from the dust on the gas
$\sim \rho_\mathrm{dust} \bm{a}_\mathrm{ext,dust}$ \citep{seligman2019non}. As
illustrated in Fig. \ref{fig:slice_balance}, the morphology of the gas density,
dust distribution and magnetic field strength are highly correlated, and the
system reaches approximately dynamical equilibrium between the gas pressure
$P_\mathrm{gas}$, the magnetic pressure $\sim |\bm{B}|^2 / 8\pi$ and the dust
ram pressure $P_\mathrm{dust}$ (estimated as the dust momentum flux across the
surface of the dust filament $\rho_\mathrm{dust} v_{z,\mathrm{dust}}\,
v_{xy,\mathrm{dust}}$). Starting from a random distribution initially, CRs
strongly react to RDI-induced magnetic field perturbations, being scattered and
transiently becoming highly clustered together with the dust before scattering
leads to their returning to a nearly random spatial distribution again at the
saturation stage (but now with a nearly-isotropic pitch angle distribution, as
we show below). Initially the CRs react coherently small-scale modes (which have
$\lambda \ll r_{L,\mathrm{CR}}$) because CRs initially have a perfectly coherent
($\delta$ function) distribution function and magnetic fields are distorted by
the RDIs; but as the CRs scatter, these local ``concentrations'' disperse.

Fig. \ref{fig:PDF} shows the probability density functions (PDFs) of densities
(top) and velocities / Lorentz factor (bottom) for gas (left), dust (middle) and
CRs (right). Density PDFs are obtained by mapping particle mass onto grids, and
velocity / Lorentz factor PDFs are weighted directly by particle masses.
Compared with gas densities which only vary by $\sim 20\%$, the PDF of grain
densities spans over two orders of magnitude and become highly non-Gaussian at
the non-linear and saturation stages, with a flat tail extending to
$\rho_\mathrm{dust} / \rho_\mathrm{dust}^0 > 10$ (similar to other pure-RDI
simulations in, e.g., \citealt{seligman2019non,hopkins2020simulating}). The
relatively uniform PDF is qualitatively maintained in the high-resolution run,
indicating that if there is a characteristic clumpiness, it's not yet being
recovered by the simulation. The CR density PDFs are qualitatively similar to
the dust density PDFs at the nonlinear stage, as the initially coherent CRs are
``dragged'' by the RDIs, while at the saturation stage when CRs become fully
scattered, the CR density PDF recovers its initial Gaussian shape (for the
reasons above). The gas velocity PDFs suggest that gas is significantly
accelerated by dust feedback up to rms velocities $\langle v_\mathrm{gas} /
c_s^0 \rangle \sim 10^{-2}$ -- $10^{-1}$. Dust grains are gently decelerated
with time, and the velocity PDFs do not reach an equilibrium state by the end of
the simulation. The CR Lorentz factor PDFs gradually broaden out with time,
indicating CRs are mildly accelerated or decelerated by $\sim 3\%$ during their
interaction with local magnetic fields (i.e., some ``diffusive re-acceleration''
effects with a non-zero CR momentum diffusion coefficient $D_{pp}$, as will be
discussed in \S\ref{sec:without_feedback}).

\subsection{Growth Rates vs.\ Linear Theory}

\begin{figure}
  \begin{centering}
    \includegraphics[width=0.49\textwidth]{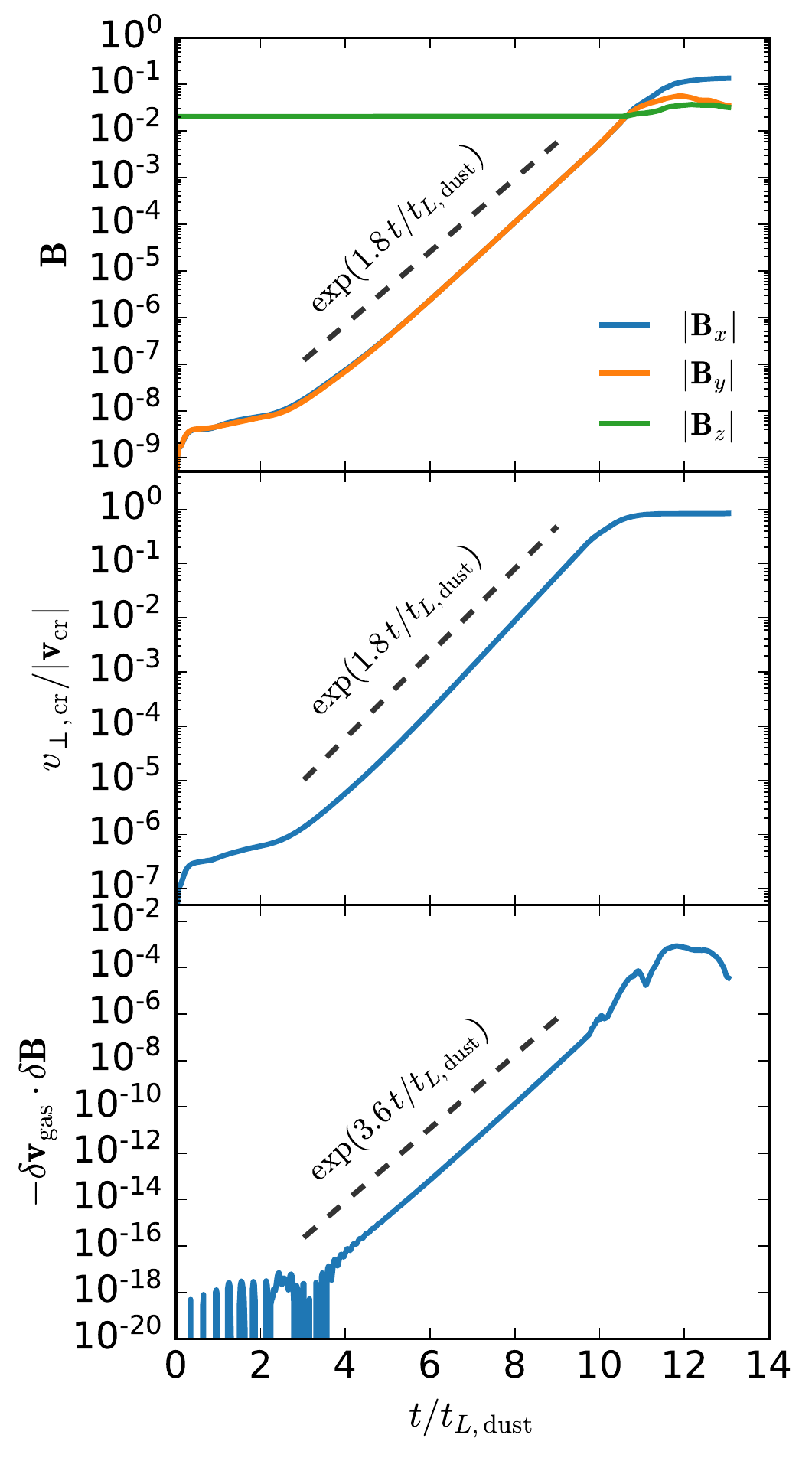}
  \end{centering}
  \vspace{-0.5cm}
  \caption{Time evolutions of averaged magnetic fields (top), the perpendicular
  component of CR velocity (middle) and the averaged negative gas cross helicity
  (bottom). Both the magnetic fields and the CR perpendicular velocity component
  follow almost the same analytically-predicted growth rate (and the squared
  growth rate for the cross helicity since it is the dot product of velocities
  and magnetic fields), indicating a strong correlation between magnetic field
  fluctuations, Alfv\'{e}n wave propagation and CR scattering.
  \label{fig:time_evolution_B_V}\vspace{-0.2cm}}
\end{figure}

\begin{figure}
  \begin{centering}
    \includegraphics[width=0.49\textwidth]{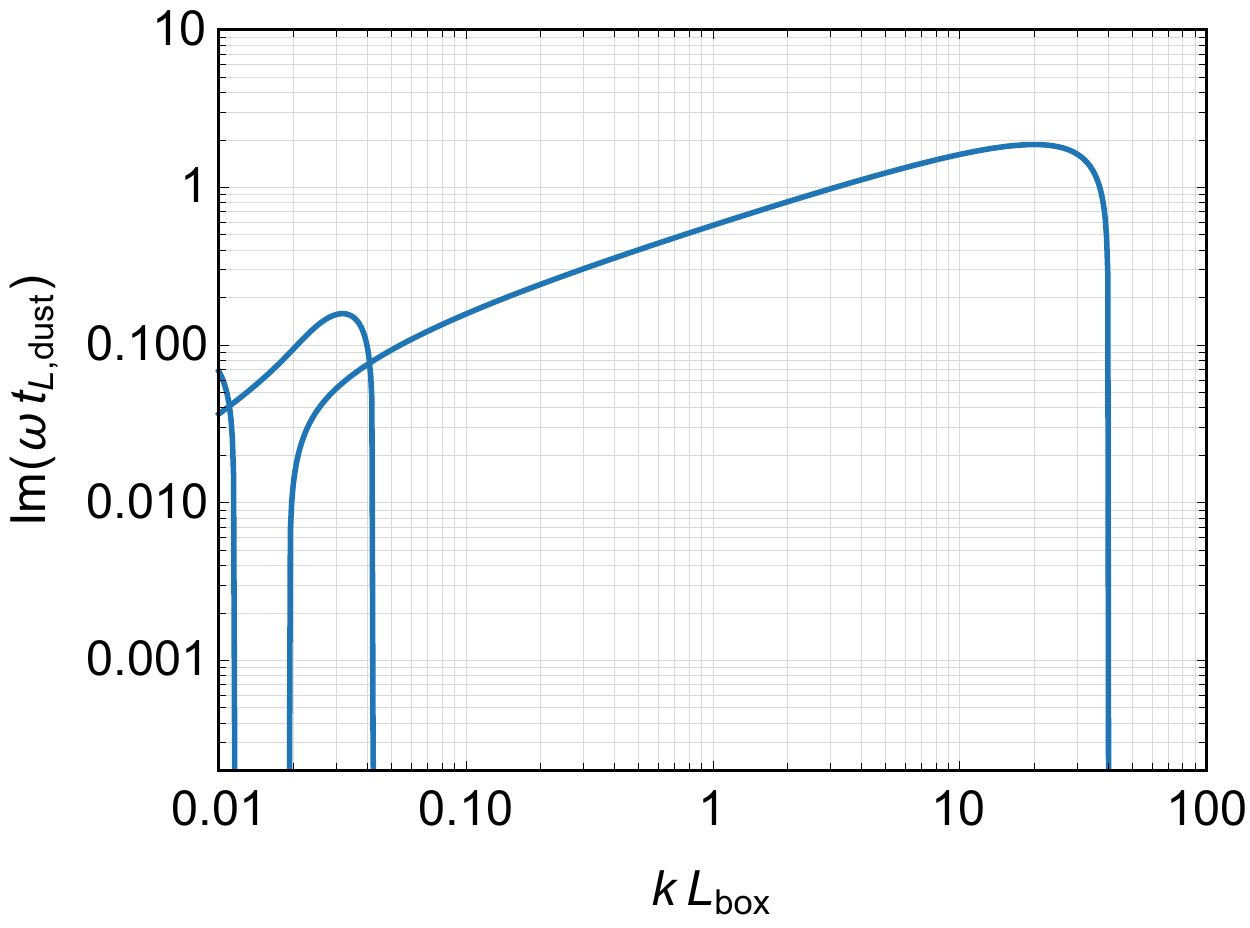}
  \end{centering}
  \vspace{-0.5cm}
  \caption{Analytical growth rates of RDI as a function of wave number
  $|\bm{k}|$, which perfectly predicts the growth rate measured directly from
  the simulation as shown in Fig. \ref{fig:time_evolution_B_V}.
  \label{fig:linear}\vspace{-0.2cm}}
\end{figure}

We next examine the amplification of magnetic fields $|\bm{B}_x|$, $|\bm{B}_y|$
and $|\bm{B}_z|$, and the growth of the CR velocity component perpendicular to
local magnetic fields $v_\mathrm{\perp,cr}$ (normalized to the magnitude of
total CR velocity $|\bm{v}_\mathrm{cr}|$). By defining the pitch angle $\theta$
as the angle between CR velocities and local magnetic field vectors, we see
$v_\mathrm{\parallel,cr} / |\bm{v}_\mathrm{cr}| = \mu$ and $v_\mathrm{\perp,cr}
/ |\bm{v}_\mathrm{cr}| = (1-\mu^2)^{1/2}$, where $\mu \equiv
\mathrm{cos}\,\theta$ is the pitch angle cosine and $v_{\parallel,\mathrm{cr}}$
the CR velocity component parallel with local magnetic fields. As shown in the
top and middle panels of Fig. \ref{fig:time_evolution_B_V}, $|\bm{B}_x|$,
$|\bm{B}_y|$ and $v_\mathrm{\perp,cr}$ grow exponentially at almost the same
growth rate until they reach saturation at $t\sim 10\,t_{L,\mathrm{dust}}$. This
indicates that, starting from highly anisotropic initial conditions with
$v_\mathrm{\perp,cr} / |\bm{v}_\mathrm{cr}| = 0$, CRs are strongly scattered by
increasingly distorted magnetic fields due to the development of the dust RDI. 

To estimate the magnitude of CR bulk drift velocity, we measure the arithmetic
mean values of the CR velocity components (the net drift velocity) over all CR
particles at the saturation stage. We find that $\left\langle
v_{\{xyz\},\mathrm{cr}}\right\rangle \sim \alpha c$ with $\alpha$ fluctuating
between $\sim 10^{-4}$ -- $10^{-2}$ with time, which is thus highly isotropic
compared with the CR initial conditions of $\langle v_{x,\mathrm{cr}}\rangle^0 =
\langle v_{y,\mathrm{cr}}\rangle^0 = 0$ and $\langle v_{z,\mathrm{cr}} \rangle^0
\sim c$. Although the magnitude of the ``residual'' drift velocity
$\sqrt{\Sigma_{xyz}^i \langle v_{i,\mathrm{cr}}\rangle^2}$ is of the same order
of magnitude with $\langle v_\mathrm{A}\rangle$, given that we sample the CR
distribution function with $\sim 10^{6}$ particles in our default simulations,
we caution that this range of $\alpha$ is precisely what we would expect from
Monte Carlo sampling noise for an intrinsically uniform pitch-angle
distribution. This sampling noise (which unfortunately converges slowly, as
$N_\mathrm{cr}^{-1/2}$) dominates the ``residual'' drift velocity here, thus we
cannot draw a firm conclusion on the CR drift speed from direct measurement.
However, there is still a way to investigate the CR drifting by Alfv\'{e}n
waves. As shown in the bottom panel of Fig. \ref{fig:time_evolution_B_V}, the
(negative) magnitude of the gas cross helicity $- \delta \bm{v}_\mathrm{gas}
\cdot \delta \bm{B}$, which is related to the asymmetry of Alfv\'{e}n waves,
remains a single sign and grows exponentially with a squared RDI growth rate
before saturating. This indicates that the propagation of the Alfv\'{e}n waves
excited by the RDIs is increasingly \emph{unidirectional} (antiparallel to the
large-scale magnetic fields in this case when the cross helicity is negative);
therefore, the CRs must at least drift at $\sim v_\mathrm{A}$ by the Alfv\'{e}n
waves all propagating in one direction.

In Fig. \ref{fig:linear}, we plot the analytically predicted growth rates for
our simulation parameters as a function of the wave number $|\bm{k}|$, for a
specific mode angle of $\hat{\bm{k}}\cdot \hat{\bm{v}}_\mathrm{dust} =
0$.\footnote{We explicitly verified that the mode angle of $\hat{\bm{k}}\cdot
\hat{\bm{v}}_\mathrm{dust} = 0$, i.e., the wave vector and dust velocities are
aligned or anti-aligned, gives the fastest growth rate, which is consistent with
the findings in \citep{seligman2019non}. Therefore, we only show the growth
rates for this specific mode angle here. For a detailed description of
calculating these growth rates, see \citet{hopkins2018ubiquitous}.} We find that
the growth rate measured from our simulation in Fig.
\ref{fig:time_evolution_B_V} replicates the analytical solution well, with the
growth rates peaking at wave numbers around $k L_\mathrm{box} \sim 10$ -- $40$.
This corresponds to a fastest growing wavelength of $2\pi/k \sim 0.16$ -- $0.6
L_\mathrm{box}$, which is somewhat consistent with the structures seen in Fig.
\ref{fig:slice} and \ref{fig:slice_y}.

\subsection{Pitch Angle Scattering \&\ Transport/Scattering Coefficients}
\label{sec:without_feedback}

\begin{figure}
  \begin{centering}
    \includegraphics[width=0.5\textwidth]{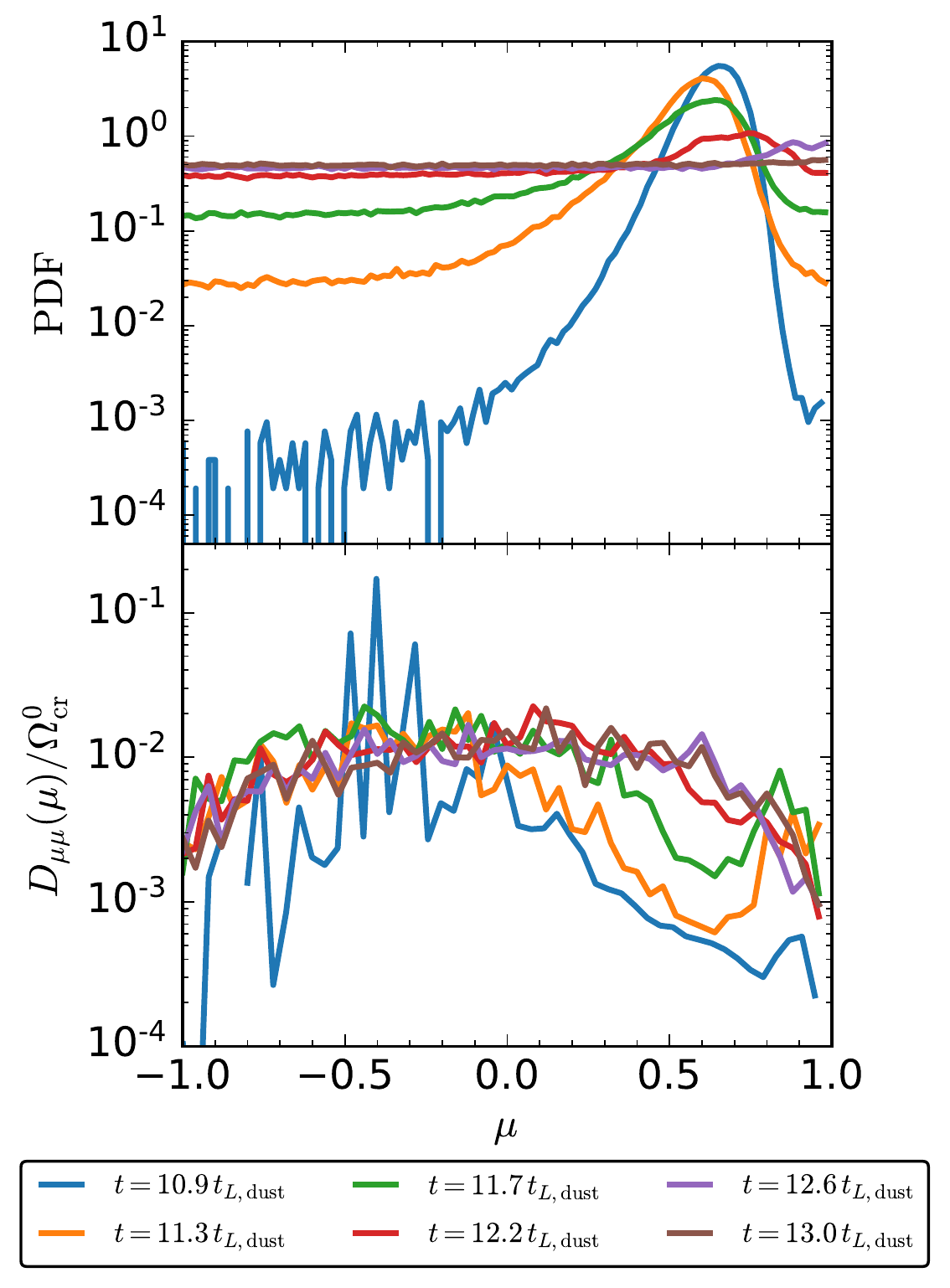}
  \end{centering}
  \vspace{-0.5cm}
  \caption{Time evolution of PDFs of the CR pitch angle cosine $\mu$ (top), and
  the CR pitch angle diffusion coefficient $D_{\mu\mu}$ (normalized by the
  initial CR gyro-frequency $\Omega_\mathrm{cr}^0$) as a function of the CR
  pitch angle cosine (bottom). At the saturation stage, CRs are fully
  isotropized with a uniform distribution of the pitch angle cosine, and the CR
  pitch angle diffusion coefficient features a flat profile around $\mu=0$,
  without encountering the $90\degree$ pitch angle problem predicted by
  quasi-linear theories. \label{fig:pitch}\vspace{-0.2cm}}
\end{figure}

Fig. \ref{fig:pitch} shows the time evolution of CR pitch angle PDFs (top) and
the pitch angle diffusion coefficients $D_{\mu\mu}$ (bottom), which is defined
as:
\begin{align}
\label{eqn:Dmm}  D_{\mu\mu} (\mu_0, t_0) = \frac{\left\langle (\mu - \mu_0)^2\right\rangle}{2 (\tilde{t} - \tilde{t}_0)} \quad \text{for $\tilde{t}-\tilde{t}_0 = \Delta \tilde{t}$ not too large,}
\end{align}
where $\mu_0$($\tilde{t}_0$) and $\mu$($\tilde{t}$) are initial and final pitch
angle cosine (time) respectively, $\Delta t$ is the integration time, and we
trace the change of pitch angles for all CR particles over a short time interval
to keep $(\mu - \mu_0)$ small, following e.g.,
\citet{beresnyak2011numerical,xu2013cosmic}. Here $\tilde{t} \equiv
(\tilde{c}/c)\,t_{\rm code}$ denotes the in-code time $t_{\rm code}$ multiplied
by the factor $\tilde{c}/c$ to correct for the reduced speed of light (RSOL)
approximation.\footnote{Specifically, as shown in \citet{ji2021accurately}, the
implementation of the RSOL in our code is mathematically equivalent to taking
the modified form of the general Vlasov equation for the CR distribution
function to be: $(c/\tilde{c})\,\partial_{t} f_\mathrm{cr} + \bm{v}_\mathrm{cr}
\cdot \nabla_{\bm{x}} f_\mathrm{cr} + \bm{F}_\mathrm{cr}\cdot  \nabla_{\bm{p}}
f_\mathrm{cr} = \partial_{t} f_\mathrm{cr}|_{\rm coll}$, i.e.\ rescaling the
time derivative of the distribution function in the simulation frame by
$\tilde{c}/c$. This ensures that once the CR distribution function reaches
steady-state, all effects of the choice of $\tilde{c}<c$ on its properties and
on the plasma vanish, but also that the CRs ``respond'' or evolve more slowly in
time by a factor $\tilde{c}/c$ to perturbations -- effectively rescaling the
units of time ``as seen by'' the CRs. This is precisely what allows us to
uniformly increase the CR timestep by the factor $c/\tilde{c}$, which is the
purpose of the RSOL, but this means in Eq.~\ref{eqn:Dmm}, we must rescale back
to the ``true'' $\Delta \tilde{t} = (\tilde{c}/c)\,\Delta t_\mathrm{code}$ to
obtain the correct ($\tilde{c}$-independent) value of $D_{\mu\mu}$. We verify
this explicitly in simulations with varied $\tilde{c}$ below.} As shown in the
top panel of Fig. \ref{fig:pitch}, the distribution of CR pitch angles becomes
uniform at the saturation stage, indicating that  CRs are nearly isotropized,
with typical $D_{\mu\mu} \sim 0.001-0.01\,\Omega_\mathrm{cr}$ (where
$\Omega_\mathrm{cr} \equiv q_\mathrm{cr}\,|\bm{B}|/(\gamma_L m_\mathrm{cr} c)$
is the usual relativistic CR gyro-frequency).

For comparison, if we assume isotropic/grey scattering with the usual
quasi-linear theory slab scattering expressions
\citep{schlickeiser:89.cr.transport.scattering.eqns}, then we would expect the
average value of $D_{\mu\mu}$ to be given by $\langle D_{\mu\mu} \rangle \sim
(3\pi/16)\,\Omega_\mathrm{cr}\,|\delta \bm{B}|^{2}/|\bm{B}|^{2}$ \citep{Zwei13},
where $\delta \bm{B}$ represents the magnitude of magnetic fluctuations on
gyro-resonant scales. We see in Fig.~\ref{fig:pitch} that our typical
$D_{\mu\mu}$ values correspond to $|\delta\bm{B}| \sim 0.1\,|\bm{B}|$, which is
roughly what we see in Figs.~\ref{fig:slice}-\ref{fig:slice_balance} (in fact we
see slightly larger overall magnetic fluctuations, but what matters here is the
gyro-resonant, parallel component, so the effective $|\delta \bm{B}|^{2}$
entering the scattering-rate expressions we would expect to be reduced by a
factor of $\sim2-3$). Thus, at least to order of magnitude or better, the
typical scattering rate  we see is consistent with quasi-linear theory
expectations. 

Examining  $D_{\mu\mu}(\mu)$, we see that the CR pitch angle diffusion
coefficients saturates at a smooth distribution as a function of $\mu$ (nearly
$\mu$-independent), which contradicts the usual prediction of
$D_\mathrm{\mu\mu}(\mu=0) \sim 0$ from quasi-linear theory (e.g.,
\citealt{skilling1971cosmic,yan2002scattering}). This prediction from
quasi-linear theory is known as the $90\degree$ pitch angle problem, since $\mu
= 0$ corresponds to an extremely short resonance wave length which contains
insufficient energy to scatter CRs away from the $90\degree$ pitch angle, and
thus CRs might be naively expected to become ``trapped'' at $\theta= 90\degree$
without being fully isotropized (e.g.,
\citealt{giacalone1999transport,felice2001cosmic}). However, in our simulation,
since the RDI-induced magnetic field perturbation $\delta B / B$ can grow up to
a few $10^{-1}$, CRs can be scattered across the $90\degree$ pitch angle and
become isotropic without any difficulty, suggesting the dust RDI is efficient in
exciting small-scale parallel Alfv{\'e}n waves and confine CRs on $\sim
\mathrm{AU}$ scales. Previous studies have suggested that resonance broadening
due to non-linear wave-particle interactions (e.g.,
\citealt{yan.lazarian.2008:cr.propagation.with.streaming,bai2019magnetohydrodynamic})
or mirror scattering (e.g., \citealt{felice2001cosmic}) might help to avoid the
$90\degree$ pitch angle problem, which may be occurring in our simulations as
well.

With this, we can further estimate the CR parallel diffusion coefficient
$\kappa_\parallel$ from our simulation in standard fashion
\citep{earl1974diffusive}:
\begin{align}
  \kappa_\parallel \approx \frac{1}{8} \int_{-1}^1 d\mu \frac{v_\mathrm{cr}^2 (1-\mu^2)^2}{D_{\mu\mu}(\mu)}.
\end{align}
Calculating this numerically in the saturation stage, we obtain
$\kappa_\parallel \sim 1.7\times10^{4}\,c_s^0 \, L_\mathrm{box} \sim
700\,c\,r_{L,\mathrm{cr}} \sim 0.7 \times10^{27}\,(|\bm{B}|/0.1\,\mathrm{\mu
G})^{-1}\,\mathrm{cm^{2}\,s^{-1}}$, i.e.\ a factor of $\sim 1000$ larger than
the Bohm limit. This value is significantly lower than typical values of $\kappa
\sim 10^{29-30}\,\mathrm{cm^2\,s^{-1}}$ inferred from Solar system measurements
or $\gamma$-ray observations of Local Group galaxies
\citep{blasi2012diffusive,amato2018cosmic,chan2019cosmic,hopkins2019but},
suggesting that dust-induced scattering near quasars or superluminous galaxies
can lead to strong CR confinement.

In Fig.~\ref{fig:PDF}, we clearly see that there is also some non-zero diffusion
in CR momentum space. In quasi-linear theory again, the effective CR
momentum-space diffusion coefficient $D_{pp}$ is trivially related (to leading
order in $\mathcal{O}(u/c)$, where $u$ represents the background plasma
velocities) to the pitch-angle-averaged $\langle D_{\mu\mu} \rangle$ as 
\begin{align}
\label{eqn:Dpp}  \langle {D}_{pp} \rangle \sim \chi \frac{p^2_\mathrm{cr} v_\mathrm{A}^2}{v_\mathrm{cr}^2}\, \langle {D}_{\mu\mu} \rangle,
\end{align}
where the factor of $\chi$ depends on the CR distribution function, and is
$\sim1/3$ for nearly isotropic CRs \citep{hopkins:m1.cr.closure}. Measuring $
\langle {D}_{pp} \rangle $ either ``directly'' (for an ensemble of CRs, as we
estimated $D_{\mu\mu}$) or  ``indirectly'' (by measuring the broadening of the
PDF of $p_\mathrm{cr}$ or $\gamma_L$ in Fig.~\ref{fig:PDF} and comparing to an
analytic diffusion solution), we confirm that the momentum-space diffusion
coefficient estimated numerically is within tens of percent of the value one
would infer from simply inserting our measured $\langle D_{\mu\mu} \rangle$ into
Eq.~\eqref{eqn:Dpp}.

\subsection{Pitch Angle Scattering with CR Feedback}
\label{sec:with_feedback}

\begin{figure}
  \begin{centering}
    \includegraphics[width=0.5\textwidth]{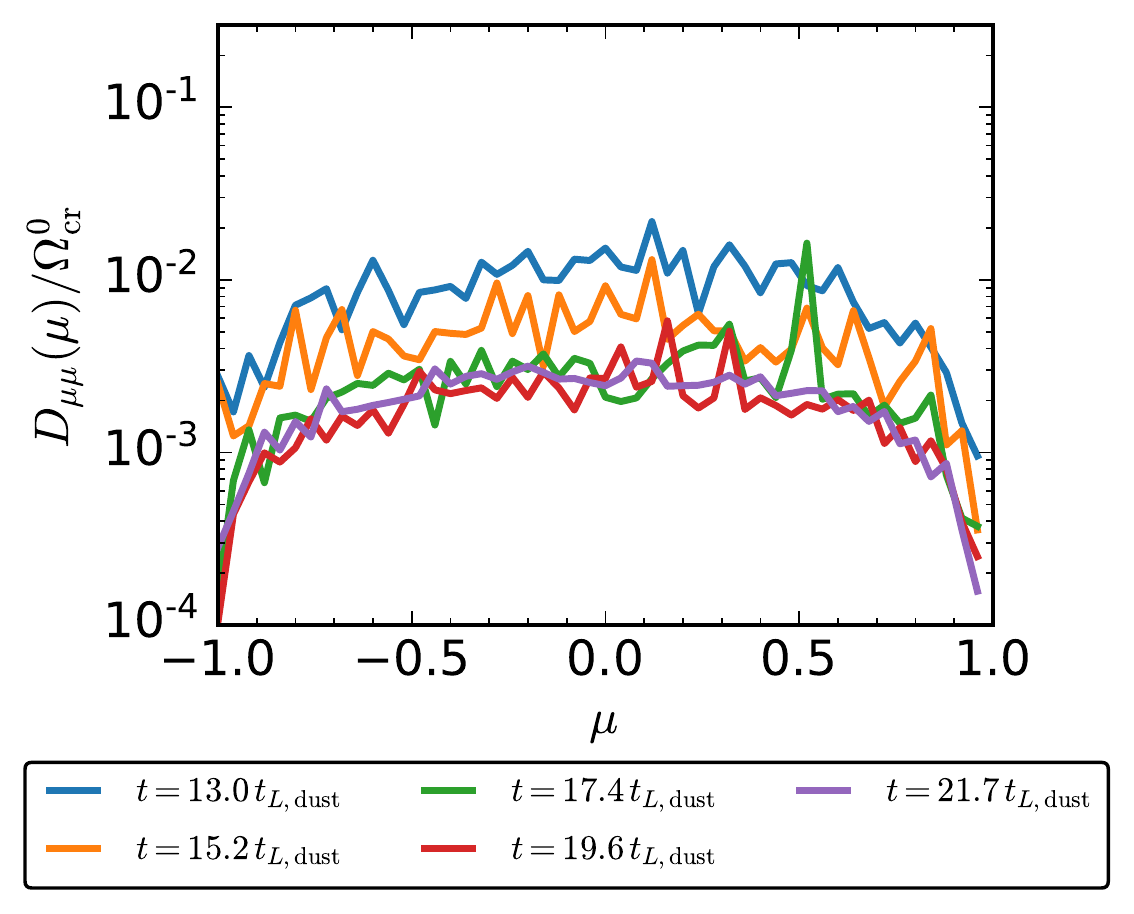}
  \end{centering}
  \vspace{-0.5cm}
  \caption{Time evolution of the CR pitch angle diffusion coefficient
  $D_{\mu\mu}$ (normalized by the initial CR gyro-frequency
  $\Omega_\mathrm{cr}^0$) as a function of the CR pitch angle cosine $\mu$, as
  the bottom panel of Fig. \ref{fig:pitch} but with CR feedback turned on. CR
  feedback lowers the CR pitch angle diffusion coefficient by roughly one order
  of magnitude, and the $90\degree$ pitch angle problem does not occur either.
  \label{fig:pitch_feedback}\vspace{-0.2cm}}
\end{figure}

We now investigate how CR feedback (back-reaction from CRs onto gas and magnetic
fields) modifies the our previous findings. After the CR population becomes
nearly isotropized at $t\sim 13t_{L,\mathrm{dust}}$ , we continue evolving our
simulation but now enable CR feedback. Since at this stage the CR drift velocity
is much less than the speed of light ($v_D \ll c$), and the CR ``initial
conditions'' are more physically realistic, the CR non-resonant instability does
not grow rapidly and dominate the simulation behavior (as it artificially would
if we began from $v_D \approx c$ with CR feedback included). Almost all of our
qualitative conclusions and the behaviors (in saturation) of gas, CR, and dust
density fields remain qualitatively similar after  turning on this back-reaction
term, but quantitatively, there is some effect. Specifically in Fig.
\ref{fig:pitch_feedback}, we plot the time evolution of the CR pitch angle
diffusion coefficient $D_{\mu\mu}$ as a function of the CR pitch angle cosine
$\mu$, after re-enabling the CR feedback. With CR feedback present, $D_{\mu\mu}$
decreases somewhat (though it conserves the functional form of
$D_{\mu\mu}(\mu)$) until saturating at a value systematically lower by a factor
$\sim 4$ compared to that without CR feedback. This in turn implies factor $\sim
4$ higher parallel diffusion coefficients. This is still more than sufficient to
keep the CRs isotropized and strongly-confined (and the $90\degree$ pitch-angle
problem still does not appear), but it is not a negligible difference.

Physically, we showed in \S~\ref{sec:without_feedback} that the scattering rates
followed approximately the quasi-linear theory expectation, $D_{\mu\mu} \propto
|\delta \bm{B}|^{2}/|\bm{B}|^{2}$. Thus a factor $\sim 4$ suppression of
$D_{\mu\mu}$ by CR back-reaction corresponds to a factor of $\sim 2$ suppression
of $\delta{\bm{B}}$ on gyro-resonant scales. Indeed, we can directly verify that
after we turn on the CR feedback the fluctuations in the magnetic field are
damped by roughly this factor. If the fluctuations are predominantly driven by
the RDIs, then this change in their saturation amplitudes is not surprising:
recall from \S~\ref{sec:parameters}, when we turn on back-reaction (i.e.\
account for finite CR pressure effects on gas), we assume a fairly large CR
pressure relative to thermal pressure, $P_\mathrm{cr} \sim
5\,P_\mathrm{thermal}$ (with both $P_\mathrm{cr}$ and $P_\mathrm{thermal}$ much
larger than magnetic pressure). Thus if we saturate as described in
\S~\ref{sec:spatial} with the perturbations driven by the RDIs (ultimately
powered by the dust ``ram pressure'' or acceleration force per unit area $P_{\rm
dust}$) compensated by gas thermal {\em plus} CR pressure, then we expect
$|\delta \bm{B}|^{2}$ to be a factor of a few lower in saturation (from the
increase in the {\em total} thermal+magnetic+CR background pressure).
Effectively, the background medium becomes ``stiffer'' against perturbation by
the dust. 

From the view point of the standard quasi-linear theory wherein one assumes
linear growth of scattering modes compensated by wave damping setting the
quasi-steady-state scattering rates, with the CR feedback turned on, the
backreaction on the RDI from the CR pressure plus the magnetic tension leads to
stronger damping, and thus less CR confinement as found in the simulations.
Quasi-linear theories also predict that the CR feedback can generate small-scale
Alfv\'en waves via the CR gyro-resonant and non-resonant instabilities, i.e.,
the CR ``self-confinement'' modes which add linearly with the RDI driven
perturbations and thus \emph{increase} the CR confinement. However, this is in
opposite to the decrease of the CR confinement in our simulations, because CRs
are highly isotropic in the saturation stage, thus the CR-excited Alfv\'en waves
and their resulting CR confinement are negligible. But admittedly, it is
plausible to imagine that if there is a continuous driving of the background CR
gradients, the growth rates of Alfv\'en waves from dust and CRs do add linearly
when they both have drifts and small densities, and we might see the enhancement
of the CR confinement, as quasi-linear self-confinement theories predicted.

\subsection{Convergence Tests}

\begin{figure*}
  \begin{centering}
    \includegraphics[width=0.983\textwidth]{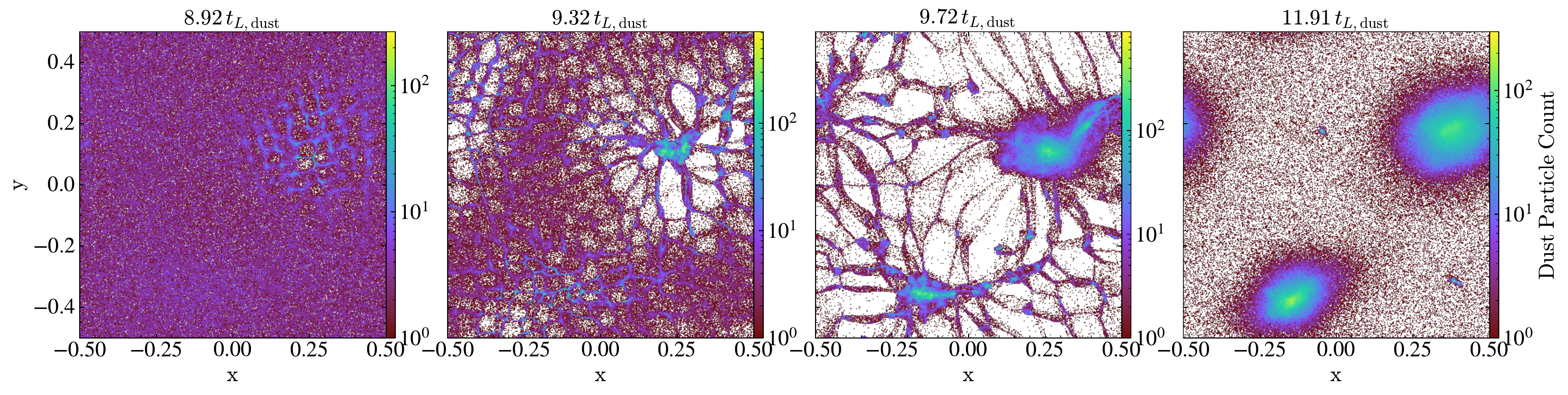}
    \includegraphics[width=\textwidth]{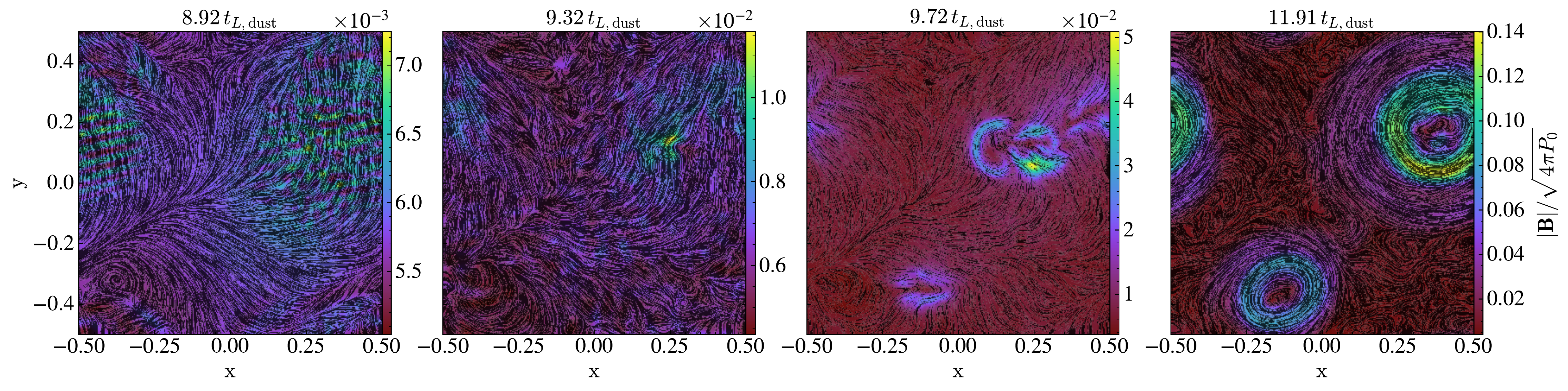}
    \includegraphics[width=0.983\textwidth]{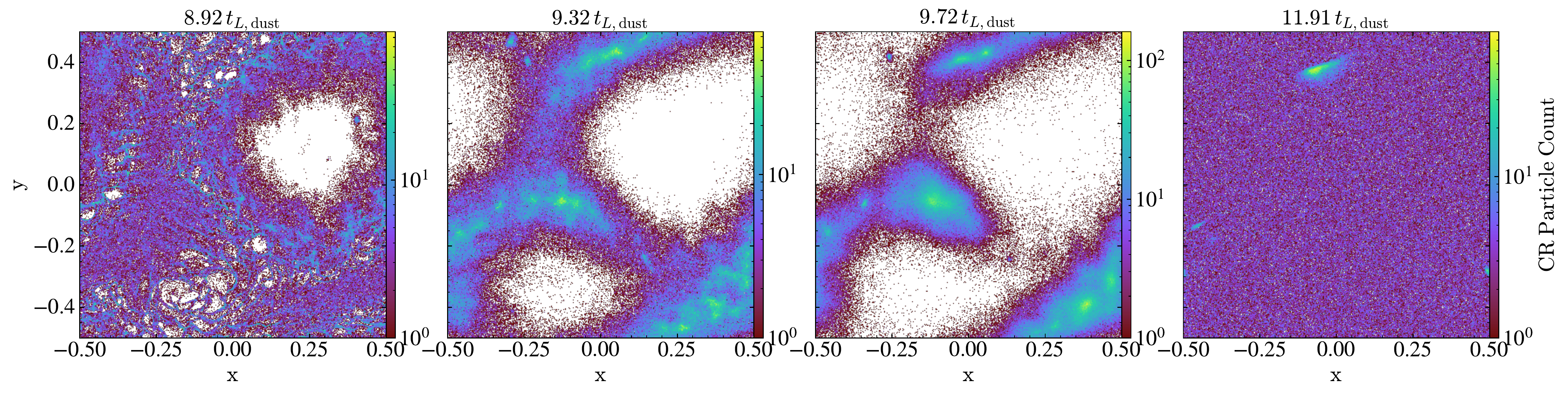}
  \end{centering}
  \caption{Plots of dust grain projections (top), magnetic field strengths
  superposed with field lines (middle) and CR particle projections (bottom), as
  Fig. \ref{fig:slice}, but from the high-resolution simulation. The morphology
  of dust grains, magnetic fields and CRs in the high-resolution run is
  qualitatively similar with those in the fiducial run, but exhibits more
  detailed small-scale structures and contains more large-scale coherent
  structures (since the domain size of the high-resolution run is effectively
  larger -- see the text for a full explanation).
  \label{fig:slice_high}\vspace{-0.2cm}}
\end{figure*}

\begin{figure}
  \begin{centering}
    \includegraphics[width=0.5\textwidth]{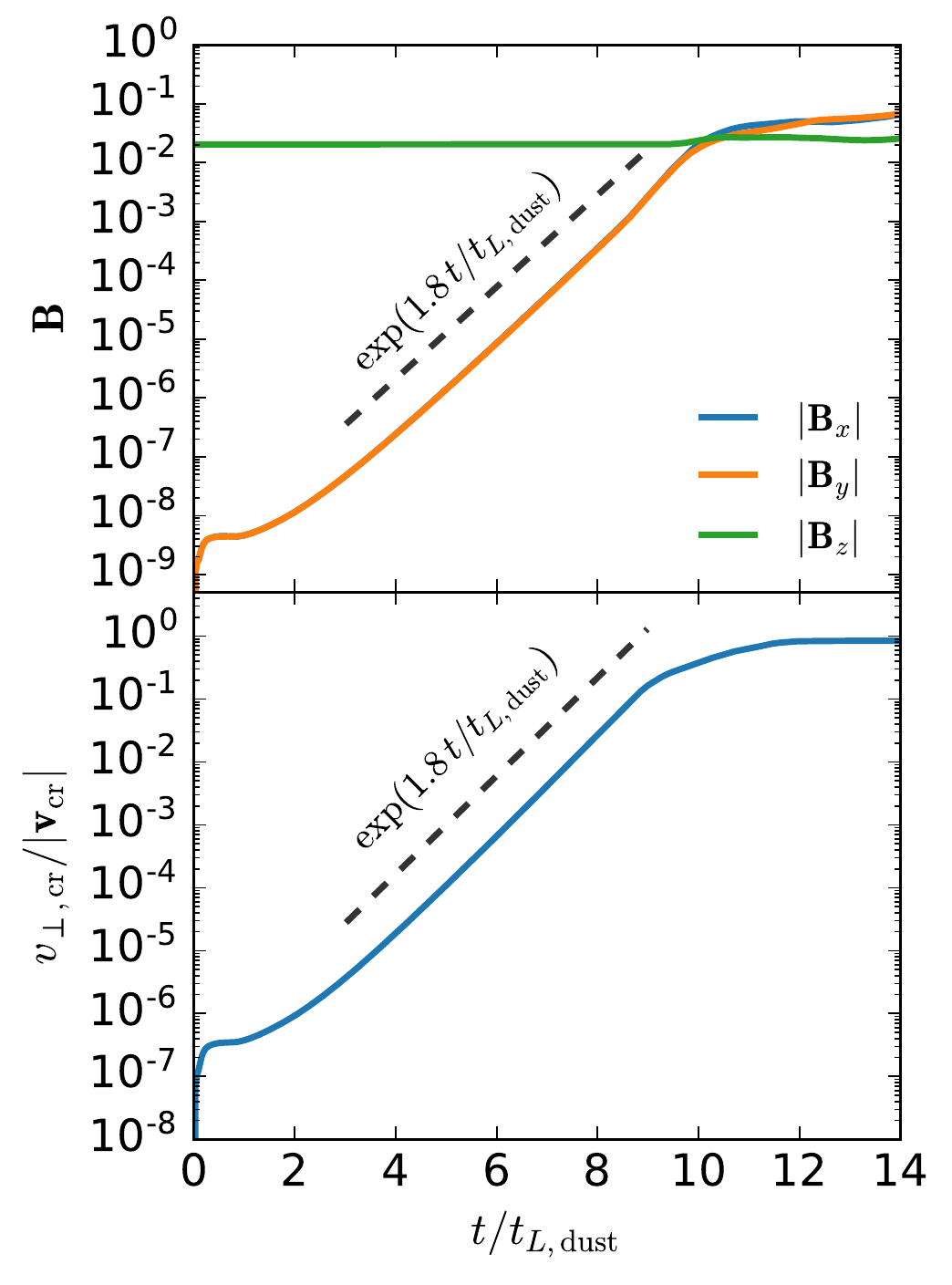}
  \end{centering}
  \vspace{-0.5cm}
  \caption{Time evolution of averaged magnetic fields (top) and the
  perpendicular component of of CR velocity, as Fig.
  \ref{fig:time_evolution_B_V}, but from the high-resolution simulation. The
  exponential growth rate is almost identical to that in the fiducial run (Fig.
  \ref{fig:time_evolution_B_V}), suggesting different resolutions do not
  significantly alter the results of our simulations.
  \label{fig:time_evolution_B_V_high}\vspace{-0.2cm}}
\end{figure}

Although previous studies have  investigated how simulations of just the RDIs
\citep{moseley2019non,seligman2019non,hopkins2020simulating} or just CRs depend
on resolution, those simulations did not combine these physics nor investigate
the same range of scales and parameters as we do here. Therefore, we consider a
``high-resolution'' simulation with $N_\mathrm{gas} = N_\mathrm{dust} =
N_\mathrm{cr} = N_\mathrm{1D}^3 = 128^3$. Note that since we enforce
$r_{L,\mathrm{dust}} / L_\mathrm{box} = \sqrt{N_\mathrm{1D}}$, the
high-resolution run with doubled 1D resolution fits a factor-of-$\sqrt{2}$
larger number of Larmor wavelengths along each side of the simulation domain,
and resolves each Larmor wavelength with $\sqrt{2}$ times more elements. 

Fig. \ref{fig:slice_high} shows projections of dust grains, slices of magnetic
field strength superposed with field lines and projections of CR particles along
the $z$-axis for the high-resolution simulation. The high-resolution run is
qualitatively similar to the low-resolution run shown in Fig. \ref{fig:slice},
but contains smaller-scale structures as expected. At the saturation stage, the
high-resolution run contains two large-scale coherent structures of dust columns
and magnetic vortices, in contrast to the single sheet-like structure in the
low-resolution run. This is likely because the high-resolution run contains more
Larmor wavelengths and thus it simulates a larger domain in real physical units.
The time evolution  of the magnetic fields and the perpendicular CR velocity in
the high-resolution run are shown in Fig. \ref{fig:time_evolution_B_V_high},
where the growth rate is almost identical to the low-resolution run. Thus the
linear growth rates of the relevant modes  are not measurably dependent on
resolution, and they agree well with analytic theory; moreover the final
properties (widths) of the PDFs are similar in both cases. Of greatest interest,
the values of $D_{\mu\mu}$ and (correspondingly) $\kappa_\parallel$ and $D_{pp}$
differ by less than $5\%$ in the high resolution run. Thus it appears that the
key results here are not sensitive to resolution. However, we caution that it is
only possible to numerically resolve a tiny fraction of the interesting dynamic
range (let alone effects like turbulent cascades or field-line wandering from
large-scale dynamics, which operate on orders-of-magnitude larger scales), so
this should be taken with some caution.

We also examine the numerical convergence with respect to the RSOL by varying
$\tilde{c}$ over a factor of $\sim5$. For standard-resolution runs with
$\tilde{c} = (0.02,\,0.05,\,0.1)\,c$, we find qualitatively identical behavior
in all properties studied here, with the numerically-estimated in-saturation
parallel diffusion coefficients $\kappa_\parallel \sim (700,\,700,\,600)\,c
r_{L,\mathrm{cr}}$, respectively -- i.e.\ nearly invariant to $\tilde{c}$. Note
there is a (very weak) hint that $\kappa_\parallel$ may decrease with increasing
$\tilde{c}$ here, perhaps owing to slightly stronger confinement at higher
$\tilde{c}$ perhaps because with lower $\tilde{c}$ the CRs response may slightly
artificially lag the RDI growth rates; but if we extrapolate to $\tilde{c}=c$ by
fitting our results, the resulting inferred $\kappa_\parallel$ is only decreased
by $\sim 40\%$, a rather small correction compared to other theoretical
uncertainties here (e.g.\ the effect of back-reaction discussed below).

\subsection{Discussion: Damping \&\ Saturation Scalings}

Owing to our limited numerical resolution and MHD-PIC assumptions (where the
plasma is treated as an MHD fluid), our simulations do not include certain
plasma processes that can also damp Alfv{\'e}n waves, such as ion-neutral or
Landau damping. Ion-neutral damping is not likely relevant  under conditions of
interest for our problem here (e.g. diffuse, warm, highly-illuminated CGM), as
the expected neutral fractions are vanishingly small. But Landau damping could
be non-negligible, in principle. If we consider e.g.\  the usual non-linear
Landau damping rate with $\Gamma \sim
(\sqrt{\pi}/4)\,c_{s}\,k\,(k^{2}_{\bot}/k^{2}_{\|})\sim
(\sqrt{\pi}/8)\,c_{s}\,k\,(|\delta\bm{B}|^{2}/|\bm{B}|^{2})$, then at the scale
where the RDI growth rate is maximized ($\sim r_{L,\mathrm{cr}}$) the implied
Landau damping rate (given the $|\delta\bm{B}| \sim 0.1\,|\bm{B}|$ we see) is
generally $\sim 10\%$ of the RDI growth rate in Fig.~\ref{fig:linear}. That
suggests it may not be negligible,  but it is also unlikely to qualitatively
change the behaviors here, if included. However, the simulations do, given the
finite resolution, have non-zero numerical dissipation which happens
(coincidentally) to be similar in magnitude to Landau damping: given a standard
numerical MHD dissipation rate in {\small  GIZMO} which  scales as $\sim \Delta
x\,c_{s}\,k^{2} \sim c_{s}\,k\,(k\,\Delta x)$ (where $\Delta x$ is the effective
grid resolution for the  MHD, and the prefactor depends on the specific
numerical problem and details of the method, see \citealt{hopkins:mhd.gizmo}),
then for $k\sim 1/r_{L,\mathrm{cr}} \sim 1/(10\,\Delta x)$ this is roughly
similar in magnitude to the physical Landau damping. Of course, other mechanisms
could in principle contribute to damping including interactions with extrinsic
turbulence \citep{farmer2004wave}, which we cannot capture owing to our
limited range of scales. Dust itself could act as a damping mechanism in some
circumstances \citep{squire:2021.dust.cr.confinement.damping} but the conditions
where this would occur are dramatically different from those here.

These limitations in physics, finite resolution, and simulation box size
preclude making detailed statements regarding the saturation mechanisms of the
RDI across a broader parameter space. However, the fact that we see roughly
isotropic $D_{\mu\mu}$ following approximately the expected quasi-linear scaling
with $|\delta \bm{B}|/|\bm{B}|$ gives us some confidence that the CR scattering
rates induced by the RDI should indeed scale with the saturation amplitude of
$\delta\bm{B}$. And even if the saturation mechanism is uncertain, some broad
conclusions are robust. For example, based on their idealized RDI-only
simulations, \citet{hopkins2020simulating} discuss two possible saturation
mechanisms, the first being balance between magnetic tension and dust ram
pressure as discussed in \S~\ref{sec:spatial}, the second being a scenario where
the crossing time of the RDI-generated modes ($\sim \delta\bm{v}_\mathrm{gas} /
\lambda$) becomes faster than the RDI growth time, producing non-linear
dissipation. Non-linear Landau damping is another potential saturation
mechanism, limiting $\delta\bm{B}$ when the damping becomes faster than linear
RDI growth rates. \citet{squire:2021.dust.cr.confinement.damping} consider yet
another saturation scenario, assuming a
\citet{kraichnan:1965.ik.aniso.turb}-like damping via self-interactions (which
gives a damping rate akin to non-linear Landau but with $c_{s}$ replaced by the
\Alf\ speed). Crucially, the ``driving'' in all of these scenarios scales with
the dust-to-gas ratio $f_\mathrm{dust-gas}$ and force/acceleration/drift
velocity on the grains (which appear in both the dust ram pressure and RDI
growth rates). This is proportional to the incident radiation flux
$F_\mathrm{rad}$. Combining these estimates for the saturation $\delta\bm{B}$
with the usual scalings for gyro radii and scattering rates we obtain, for {\em
any} of the saturation scenarios above, a scattering rate which scales
dimensionally as $D_{\mu\mu} \propto
f_\mathrm{dust-gas}\,F_\mathrm{rad}^{0.7-1.5}\,|\bm{B}|^{-(0.8-3)}$ (with a
weaker residual dependence on gas density and/or temperature, and the exact
power-law scaling depending on the saturation model). In other words, there is a
qualitatively robust prediction that with lower-dust-to-gas-ratios and/or
incident radiative fluxes and/or stronger magnetic fields, the confinement of
CRs by dust become weaker. All else equal, in order for the scattering rate from
the dust RDIs to drop to below that inferred for Milky Way ISM gas (and thus
become relatively unimportant), the factor $\sim
f_\mathrm{dust-gas}\,F_\mathrm{rad}$ would need to be $\sim 1000$ times smaller
than the value we assume to motivate our tests. So for there to be ``too little
dust,'' the metallicity or dust-to-metals ratio would need to be $1000$ times
lower than ISM values (which seems unlikely at least at low cosmological
redshifts, even in the CGM, given the observations reviewed in
\S~\ref{sec:intro}, which suggest this factor is perhaps something like $\sim
10$ times lower than in the ISM). But more plausibly, the incident flux could
easily be $1000$ times smaller, if for example the galaxy is a typical Milky
Way-like or smaller dwarf galaxy with a star formation rate of $\ll
10\,\mathrm{M_{\odot}\,yr^{-1}}$ and has negligible AGN luminosity (i.e.\ galaxy
luminosity $\lesssim 10^{10}\,\mathrm{L_{\odot}}$) or the dust is at $\gg
100\,$kpc from the host. The magnetic field dependence is also interesting: it
suggests these instabilities and ensuing confinement would be easier to excite
in the distant CGM or IGM (where nano-Gauss  fields are expected), but may be
suppressed in denser, more highly-ionized super-bubbles near to galaxies.

\section{Conclusions}
\label{sec:conclusions}

In this paper, we investigate the impact of the dust RDIs on cosmic ray
scattering, by performing the first numerical simulations of MHD-dust-CR
interactions, where the charged dust and CR gyro-radii on $\sim\mathrm{AU}$
scales are fully resolved. Since this is a first study, we consider just one
special case, where we might anticipate efficient dust-induced CR confinement,
as compared to more typical Solar-neighborhood-like ISM conditions. We focus on
\emph{one} regime of the RDIs, specifically conditions where the ``cosmic
ray-like'' RDIs are rapidly growing and can produce ``diffusive'' dust behavior
\citep[see][]{hopkins2020simulating}, which as speculated in
\citet{squire2020impact} could in turn lead to dust-induced CR confinement. This
type of RDI requires particular conditions to dominate: $v_\mathrm{dust}\gg
v_\mathrm{A}$, plasma $\beta \gg 1$, low gas density and high grain charge, so
the dust drag force is substantially subdominant to the Lorentz force. These
conditions could arise in, e.g., the CGM around quasars or luminous galaxies. We
suspect that different RDIs might have different effects on CR scattering, which
is a subject for future study. Under these conditions, we find that small-scale
parallel Alfv{\'e}n waves excited by the RDIs efficiently scatter CRs and
significantly enhance CR confinement. Therefore, dust-induced CR scattering can
potentially provide a strong CR feedback mechanism on $\sim \mathrm{AU}$ scales.

We first explore the ultra-low CR density limit by ignoring CR feedback to the
gas. Dust quickly becomes unstable to the RDIs and forms high-density columns
and sheets, until growth saturates around $\sim 10$ times the dust Larmor time
$t_{L,\mathrm{dust}}$. The density and velocity PDFs of both gas and dust grains
show strong asymmetry, where the dust density spans over two orders of
magnitude, while the gas is only slightly compressible with $\sim 10\%$ density
fluctuations. Perpendicular magnetic field components are exponentially
amplified by the RDI, with the growth rates predicted by analytical solutions.
Initially perfectly streaming CRs are strongly scattered by magnetic field
fluctuations, with the growth rate of the perpendicular CR velocity component
(to local magnetic fields, $v_{\perp,\mathrm{cr}}/|\bm{v}_\mathrm{cr}|$, or
$\sqrt{1-\mu^2}$) equaling the growth rate of magnetic fields. At the saturation
stage, the CRs are isotropized with a near-uniform distribution of the pitch
angle cosine $\mu$, and the CR pitch angle diffusion coefficient
$D_{\mu\mu}(\mu)$ is nearly independent of pitch-angle $\mu$ (in particular
around $\mu=0$). There is no $90\degree$ pitch angle problem in our simulations.
The scattering rate is in order-of-magnitude agreement with the usual
quasi-linear theory expectation $D_{\mu\mu} \sim
\Omega_\mathrm{cr}\,|\delta\bm{B}|^{2}/|\bm{B}|^{2}$, with the large
$|\delta\bm{B}|^{2}/|\bm{B}|^{2} \sim 10^{-3}-10^{-2}$ on gyro-resonant scales
driven by the dust RDIs (in part because the dust has broadly similar gyro-radii
to the CRs, under these types of conditions). The numerically-calculated CR
parallel diffusion coefficient is $\sim 500-1000$ times the Bohm value:
sufficient for strong confinement of the CRs.

When the system reaches saturation and CRs become close to isotropic (with the
CR drift velocity $v_D \ll \tilde{c}$), we turn on CR feedback to the gas and
study its consequences. We find that with CR feedback, the CR pitch angle
diffusion coefficient $D_{\mu\mu}$ decreases by a factor of $\sim 4$ (and thus
this slightly reduces CR confinement). This owes to the fact that the large
assumed CR pressure (several times larger than thermal+magnetic) suppresses the
saturation amplitude of the magnetic field fluctuations $\delta \bm{B}$ induced
by the RDIs by a modest factor $\sim 2$, and the scattering modes are thus
damped more by backreaction on the RDIs from the CR pressure, in addition to the
magnetic tension. Since CRs in the saturation stage are highly isotropic, the
quasi-linear self-confinement theory which predicts \emph{higher} scattering
rates due to CR-excited Alfv\'en waves is \emph{not} applicable here, unless
there exists a continuous driving of CR gradients.

We finally stress that several caveats apply to our study. (1) As a first
experiment, we picked one particular initial condition to investigate an
interesting case,  which is plausible for some conditions as noted above but
should not be considered typical everywhere. There exist a variety of RDIs which
can have totally different behaviors in different circumstances
\citep{hopkins2020simulating} and indeed, under some conditions dust might even
have the opposite effect, acting as a wave-damping mechanism and reducing the
confinement of CRs \citep{squire2020impact}. (2) Although we perform a small
resolution study, our simulations are still limited in resolution and dynamical
range. Even for a single grain size, the RDIs are unstable at all spatial
wavelengths, so it is impossible to encompass their complete dynamic range (let
alone global scales of structure or extrinsic turbulence, which are vastly
larger than our box sizes).  
(3) We do not include any explicit wave-damping processes. While we do not
expect appreciable ion-neutral damping in environments of interest (as the
neutral fractions are negligible), Landau damping could be important, as could
damping from a turbulent cascade, and these could reduce the efficacy of
confinement. However at least for the extreme parameters considered here, it is
unlikely these would significantly reduce the scattering rates. (4) Our periodic
boxes neglect large-scale ($\gg L_\mathrm{box}$) CR pressure gradients ($\nabla
P_\mathrm{cr}$) which act as a source/driving term for super-Alfv{\'e}nic CR
drift. (5) For simplicity, we consider only one grain size+charge and one CR
energy+species, rather than a full spectrum of grain sizes and CR energies. In
future work it will be particularly interesting to see how a full spectrum of
both modifies the dynamics here, as a broad range of gyro-radii overlap and
different gyro-resonant modes can interact non-linearly and even linearly (when
CRs+dust+gas are all combined), via their back-reaction on the gas.

With these limitations in mind, we consider this study to be a \emph{proof of
concept}, showing that CR dust-gas interactions might indeed be very important
in some astrophysical conditions. For instance, this mechanism might be able to
resolve some of the incompatibility between CR confinement models and
observations by preventing CR ``runaway'' \citep{hopkins2021standard}, at least
under certain conditions investigated in this study. Considerable work remains
to map out the parameter space, include additional physics, and understand the
macroscopic consequences of confining CR-dust interactions (for either CRs
themselves or for galaxy/CGM evolution). Nevertheless, the simulations here
clearly argue that these effects are worth studying in detail.

\dataavailability{The data supporting the plots within this article are available on reasonable request to the corresponding author. A public version of the GIZMO code is available at \gizmourl.}

\acknowledgments 
SJ thanks E. Quataert for helpful discussions, and the referee for constructive
comments which improve this manuscript. SJ is supported by a Sherman Fairchild
Fellowship from Caltech, the Natural Science Foundation of China (grants
12133008, 12192220, and 12192223) and the science research grants from the China
Manned Space Project (No. CMS-CSST-2021-B02). Support for JS  was provided by
Rutherford Discovery Fellowship RDF-U001804 and Marsden Fund grant UOO1727,
which are managed through the Royal Society Te Ap\=arangi. Support for PFH was
provided by NSF Research Grants 1911233 \&\ 20009234, NSF CAREER grant 1455342,
NASA grants 80NSSC18K0562, HST-AR-15800.001-A, JPL 1589742. Numerical
calculations were run on the Caltech compute cluster ``Wheeler,'' allocations
FTA-Hopkins/AST20016 supported by the NSF and TACC, and NASA HEC SMD-16-7592. We
have made use of NASA's Astrophysics Data System. Data analysis and
visualization are made with {\small Python 3}, and its packages including
{\small NumPy} \citep{van2011numpy}, {\small SciPy} \citep{oliphant2007python},
{\small Matplotlib} \citep{hunter2007matplotlib}, and the {\small yt}
astrophysics analysis software suite \citep{turk2010yt}.\\

\bibliography{ms_extracted}

\begin{thebibliography}{}
\makeatletter
\relax
\def\mn@urlcharsother{\let\do\@makeother \do\$\do\&\do\#\do\^\do\_\do\%\do\~}
\def\mn@doi{\begingroup\mn@urlcharsother \@ifnextchar [ {\mn@doi@}
  {\mn@doi@[]}}
\def\mn@doi@[#1]#2{\def\@tempa{#1}\ifx\@tempa\@empty \href
  {http://dx.doi.org/#2} {doi:#2}\else \href {http://dx.doi.org/#2} {#1}\fi
  \endgroup}
\def\mn@eprint#1#2{\mn@eprint@#1:#2::\@nil}
\def\mn@eprint@arXiv#1{\href {http://arxiv.org/abs/#1} {{\tt arXiv:#1}}}
\def\mn@eprint@dblp#1{\href {http://dblp.uni-trier.de/rec/bibtex/#1.xml}
  {dblp:#1}}
\def\mn@eprint@#1:#2:#3:#4\@nil{\def\@tempa {#1}\def\@tempb {#2}\def\@tempc
  {#3}\ifx \@tempc \@empty \let \@tempc \@tempb \let \@tempb \@tempa \fi \ifx
  \@tempb \@empty \def\@tempb {arXiv}\fi \@ifundefined
  {mn@eprint@\@tempb}{\@tempb:\@tempc}{\expandafter \expandafter \csname
  mn@eprint@\@tempb\endcsname \expandafter{\@tempc}}}

\bibitem[\protect\citeauthoryear{Amato \& Blasi}{Amato \&
  Blasi}{2018}]{amato2018cosmic}
Amato E.,  Blasi P.,  2018, Advances in Space Research, 62, 2731

\bibitem[\protect\citeauthoryear{{Bai} \& {Stone}}{{Bai} \&
  {Stone}}{2010}]{bai:2010.grain.streaming.vs.diskparams}
{Bai} X.-N.,  {Stone} J.~M.,  2010, \mn@doi [\apjl]
  {10.1088/2041-8205/722/2/L220}, \href
  {http://adsabs.harvard.edu/abs/2010ApJ...722L.220B} {722, L220}

\bibitem[\protect\citeauthoryear{Bai, Caprioli, Sironi  \& Spitkovsky}{Bai
  et~al.}{2015}]{bai2015magnetohydrodynamic}
Bai X.-N.,  Caprioli D.,  Sironi L.,   Spitkovsky A.,  2015, The Astrophysical
  Journal, 809, 55

\bibitem[\protect\citeauthoryear{Bai, Ostriker, Plotnikov  \& Stone}{Bai
  et~al.}{2019}]{bai2019magnetohydrodynamic}
Bai X.-N.,  Ostriker E.~C.,  Plotnikov I.,   Stone J.~M.,  2019, The
  Astrophysical Journal, 876, 60

\bibitem[\protect\citeauthoryear{Bell}{Bell}{1978}]{bell1978acceleration}
Bell A.,  1978, Monthly Notices of the Royal Astronomical Society, 182, 147

\bibitem[\protect\citeauthoryear{{Bell}}{{Bell}}{2004}]{bell.2004.cosmic.rays}
{Bell} A.~R.,  2004, \mn@doi [\mnras] {10.1111/j.1365-2966.2004.08097.x}, \href
  {http://adsabs.harvard.edu/abs/2004MNRAS.353..550B} {353, 550}

\bibitem[\protect\citeauthoryear{Beresnyak, Yan  \& Lazarian}{Beresnyak
  et~al.}{2011}]{beresnyak2011numerical}
Beresnyak A.,  Yan H.,   Lazarian A.,  2011, The Astrophysical Journal, 728, 60

\bibitem[\protect\citeauthoryear{Blasi \& Amato}{Blasi \&
  Amato}{2012}]{blasi2012diffusive}
Blasi P.,  Amato E.,  2012, Journal of Cosmology and Astroparticle Physics,
  2012, 010

\bibitem[\protect\citeauthoryear{Buck, Pfrommer, Pakmor, Grand  \&
  Springel}{Buck et~al.}{2020}]{buck2020effects}
Buck T.,  Pfrommer C.,  Pakmor R.,  Grand R.~J.,   Springel V.,  2020, Monthly
  Notices of the Royal Astronomical Society, 497, 1712

\bibitem[\protect\citeauthoryear{Butsky \& Quinn}{Butsky \&
  Quinn}{2018}]{butsky2018role}
Butsky I.~S.,  Quinn T.~R.,  2018, The Astrophysical Journal, 868, 108

\bibitem[\protect\citeauthoryear{{Carballido}, {Stone}  \&
  {Turner}}{{Carballido}
  et~al.}{2008}]{carballido:2008.grain.streaming.instab.sims}
{Carballido} A.,  {Stone} J.~M.,   {Turner} N.~J.,  2008, \mn@doi [\mnras]
  {10.1111/j.1365-2966.2008.13014.x}, \href
  {http://adsabs.harvard.edu/abs/2008MNRAS.386..145C} {386, 145}

\bibitem[\protect\citeauthoryear{Chan, Kere{\v{s}}, Hopkins, Quataert, Su,
  Hayward  \& Faucher-Gigu{\`e}re}{Chan et~al.}{2019}]{chan2019cosmic}
Chan T.,  Kere{\v{s}} D.,  Hopkins P.,  Quataert E.,  Su K.,  Hayward C.,
  Faucher-Gigu{\`e}re C.,  2019, Monthly Notices of the Royal Astronomical
  Society, 488, 3716

\bibitem[\protect\citeauthoryear{{Deng}, {Mayer}, {Latter}, {Hopkins}  \&
  {Bai}}{{Deng} et~al.}{2019}]{deng:2019.mri.turb.sims.gizmo.methods}
{Deng} H.,  {Mayer} L.,  {Latter} H.,  {Hopkins} P.~F.,   {Bai} X.-N.,  2019,
  \mn@doi [\apjs] {10.3847/1538-4365/ab0957}, \href
  {https://ui.adsabs.harvard.edu/abs/2019ApJS..241...26D} {241, 26}

\bibitem[\protect\citeauthoryear{{Draine} \& {Sutin}}{{Draine} \&
  {Sutin}}{1987}]{draine:1987.grain.charging}
{Draine} B.~T.,  {Sutin} B.,  1987, \mn@doi [\apj] {10.1086/165596}, \href
  {http://adsabs.harvard.edu/abs/1987ApJ...320..803D} {320, 803}

\bibitem[\protect\citeauthoryear{Earl}{Earl}{1974}]{earl1974diffusive}
Earl J.,  1974, The Astrophysical Journal, 193, 231

\bibitem[\protect\citeauthoryear{{Evoli}, {Gaggero}, {Vittino}, {Di Bernardo},
  {Di Mauro}, {Ligorini}, {Ullio}  \& {Grasso}}{{Evoli}
  et~al.}{2017}]{evoli:dragon2.cr.prop}
{Evoli} C.,  {Gaggero} D.,  {Vittino} A.,  {Di Bernardo} G.,  {Di Mauro} M.,
  {Ligorini} A.,  {Ullio} P.,   {Grasso} D.,  2017, \mn@doi [Journal of
  Cosmology and Astroparticle Physics] {10.1088/1475-7516/2017/02/015}, \href
  {http://adsabs.harvard.edu/abs/2017JCAP...02..015E} {2, 015}

\bibitem[\protect\citeauthoryear{Farber, Ruszkowski, Yang  \& Zweibel}{Farber
  et~al.}{2018}]{farber2018impact}
Farber R.,  Ruszkowski M.,  Yang H.-Y.,   Zweibel E.,  2018, The Astrophysical
  Journal, 856, 112

\bibitem[\protect\citeauthoryear{Farmer \& Goldreich}{Farmer \&
  Goldreich}{2004}]{farmer2004wave}
Farmer A.~J.,  Goldreich P.,  2004, The Astrophysical Journal, 604, 671

\bibitem[\protect\citeauthoryear{Felice \& Kulsrud}{Felice \&
  Kulsrud}{2001}]{felice2001cosmic}
Felice G.~M.,  Kulsrud R.,  2001, The Astrophysical Journal, 553, 198

\bibitem[\protect\citeauthoryear{Giacalone \& Jokipii}{Giacalone \&
  Jokipii}{1999}]{giacalone1999transport}
Giacalone J.,  Jokipii J.,  1999, The Astrophysical Journal, 520, 204

\bibitem[\protect\citeauthoryear{{Grudi{\'c}}, {Guszejnov}, {Hopkins}, {Offner}
   \& {Faucher-Gigu{\`e}re}}{{Grudi{\'c}}
  et~al.}{2020}]{grudic:starforge.methods}
{Grudi{\'c}} M.~Y.,  {Guszejnov} D.,  {Hopkins} P.~F.,  {Offner} S. S.~R.,
  {Faucher-Gigu{\`e}re} C.-A.,  2020, MNRAS, submitted, arXiv:2010.11254, \href
  {https://ui.adsabs.harvard.edu/abs/2020arXiv201011254G} {p. arXiv:2010.11254}

\bibitem[\protect\citeauthoryear{{Guo} \& {Oh}}{{Guo} \&
  {Oh}}{2008}]{guo.oh:cosmic.rays}
{Guo} F.,  {Oh} S.~P.,  2008, \mn@doi [\mnras]
  {10.1111/j.1365-2966.2007.12692.x}, \href
  {http://adsabs.harvard.edu/abs/2008MNRAS.384..251G} {384, 251}

\bibitem[\protect\citeauthoryear{Haggerty, Caprioli  \& Zweibel}{Haggerty
  et~al.}{2019}]{haggerty2019hybrid}
Haggerty C.,  Caprioli D.,   Zweibel E.,  2019, arXiv preprint arXiv:1909.06346

\bibitem[\protect\citeauthoryear{Hennawi et~al.,}{Hennawi
  et~al.}{2010}]{hennawi2010binary}
Hennawi J.~F.,  et~al., 2010, The Astrophysical Journal, 719, 1672

\bibitem[\protect\citeauthoryear{{Holcomb} \& {Spitkovsky}}{{Holcomb} \&
  {Spitkovsky}}{2019}]{holcolmb.spitkovsky:saturation.gri.sims}
{Holcomb} C.,  {Spitkovsky} A.,  2019, \mn@doi [\apj]
  {10.3847/1538-4357/ab328a}, \href
  {https://ui.adsabs.harvard.edu/abs/2019ApJ...882....3H} {882, 3}

\bibitem[\protect\citeauthoryear{{Hopkins}}{{Hopkins}}{2015}]{hopkins:gizmo}
{Hopkins} P.~F.,  2015, \mn@doi [\mnras] {10.1093/mnras/stv195}, \href
  {http://adsabs.harvard.edu/abs/2015MNRAS.450...53H} {450, 53}

\bibitem[\protect\citeauthoryear{{Hopkins}}{{Hopkins}}{2016}]{hopkins:cg.mhd.gizmo}
{Hopkins} P.~F.,  2016, \mn@doi [\mnras] {10.1093/mnras/stw1578}, \href
  {http://adsabs.harvard.edu/abs/2016MNRAS.462..576H} {462, 576}

\bibitem[\protect\citeauthoryear{{Hopkins}}{{Hopkins}}{2017}]{hopkins:gizmo.diffusion}
{Hopkins} P.~F.,  2017, \mn@doi [\mnras] {10.1093/mnras/stw3306}, \href
  {http://adsabs.harvard.edu/abs/2017MNRAS.466.3387H} {466, 3387}

\bibitem[\protect\citeauthoryear{{Hopkins} \& {Lee}}{{Hopkins} \&
  {Lee}}{2016}]{hopkins.2016:dust.gas.molecular.cloud.dynamics.sims}
{Hopkins} P.~F.,  {Lee} H.,  2016, \mn@doi [\mnras] {10.1093/mnras/stv2745},
  \href {http://adsabs.harvard.edu/abs/2016MNRAS.456.4174H} {456, 4174}

\bibitem[\protect\citeauthoryear{{Hopkins} \& {Raives}}{{Hopkins} \&
  {Raives}}{2016}]{hopkins:mhd.gizmo}
{Hopkins} P.~F.,  {Raives} M.~J.,  2016, \mn@doi [\mnras]
  {10.1093/mnras/stv2180}, \href
  {http://adsabs.harvard.edu/abs/2016MNRAS.455...51H} {455, 51}

\bibitem[\protect\citeauthoryear{Hopkins \& Squire}{Hopkins \&
  Squire}{2018}]{hopkins2018ubiquitous}
Hopkins P.~F.,  Squire J.,  2018, Monthly Notices of the Royal Astronomical
  Society, 479, 4681

\bibitem[\protect\citeauthoryear{Hopkins et~al.,}{Hopkins
  et~al.}{2019}]{hopkins2019but}
Hopkins P.~F.,  et~al., 2019, arXiv preprint arXiv:1905.04321

\bibitem[\protect\citeauthoryear{Hopkins, Squire, Chan, Quataert, Ji, Keres  \&
  Faucher-Gigu{\`e}re}{Hopkins et~al.}{2020a}]{hopkins2020testing}
Hopkins P.~F.,  Squire J.,  Chan T.,  Quataert E.,  Ji S.,  Keres D.,
  Faucher-Gigu{\`e}re C.-A.,  2020a, arXiv preprint arXiv:2002.06211

\bibitem[\protect\citeauthoryear{Hopkins, Squire  \& Seligman}{Hopkins
  et~al.}{2020b}]{hopkins2020simulating}
Hopkins P.~F.,  Squire J.,   Seligman D.,  2020b, Monthly Notices of the Royal
  Astronomical Society, 496, 2123

\bibitem[\protect\citeauthoryear{Hopkins, Squire, Butsky  \& Ji}{Hopkins
  et~al.}{2021a}]{hopkins2021standard}
Hopkins P.~F.,  Squire J.,  Butsky I.~S.,   Ji S.,  2021a, arXiv preprint
  arXiv:2112.02153

\bibitem[\protect\citeauthoryear{{Hopkins}, {Squire}  \& {Butsky}}{{Hopkins}
  et~al.}{2021b}]{hopkins:m1.cr.closure}
{Hopkins} P.~F.,  {Squire} J.,   {Butsky} I.~S.,  2021b, \mnras, submitted,
  arXiv:2103.10443, \href
  {https://ui.adsabs.harvard.edu/abs/2021arXiv210310443H} {p. arXiv:2103.10443}

\bibitem[\protect\citeauthoryear{Hunter}{Hunter}{2007}]{hunter2007matplotlib}
Hunter J.~D.,  2007, Computing in science \& engineering, 9, 90

\bibitem[\protect\citeauthoryear{Ji \& Hopkins}{Ji \&
  Hopkins}{2021}]{ji2021accurately}
Ji S.,  Hopkins P.~F.,  2021, Accurately Incorporating a Reduced-Speed-of-Light
  in Magnetohydrodynamic-Particle-in-Cell Simulations (\mn@eprint {arXiv}
  {2111.14704})

\bibitem[\protect\citeauthoryear{Ji et~al.,}{Ji
  et~al.}{2020}]{ji2020properties}
Ji S.,  et~al., 2020, Monthly Notices of the Royal Astronomical Society, 496,
  4221

\bibitem[\protect\citeauthoryear{Ji, Kere{\v{s}}, Chan, Stern, Hummels,
  Hopkins, Quataert  \& Faucher-Gigu{\`e}re}{Ji et~al.}{2021}]{ji2021virial}
Ji S.,  Kere{\v{s}} D.,  Chan T.,  Stern J.,  Hummels C.~B.,  Hopkins P.~F.,
  Quataert E.,   Faucher-Gigu{\`e}re C.-A.,  2021, Monthly Notices of the Royal
  Astronomical Society, 505, 259

\bibitem[\protect\citeauthoryear{{J{\'o}hannesson} et~al.,}{{J{\'o}hannesson}
  et~al.}{2016}]{2016ApJ...824...16J}
{J{\'o}hannesson} G.,  et~al., 2016, \mn@doi [\apj]
  {10.3847/0004-637X/824/1/16}, \href
  {http://adsabs.harvard.edu/abs/2016ApJ...824...16J} {824, 16}

\bibitem[\protect\citeauthoryear{{Johansen}, {Youdin}  \& {Mac Low}}{{Johansen}
  et~al.}{2009}]{johansen:2009.particle.clumping.metallicity.dependence}
{Johansen} A.,  {Youdin} A.,   {Mac Low} M.-M.,  2009, \mn@doi [\apjl]
  {10.1088/0004-637X/704/2/L75}, \href
  {http://adsabs.harvard.edu/abs/2009ApJ...704L..75J} {704, L75}

\bibitem[\protect\citeauthoryear{Jokipii}{Jokipii}{1966}]{jokipii1966cosmic}
Jokipii J.~R.,  1966, The Astrophysical Journal, 146, 480

\bibitem[\protect\citeauthoryear{{Kraichnan}}{{Kraichnan}}{1965}]{kraichnan:1965.ik.aniso.turb}
{Kraichnan} R.~H.,  1965, \mn@doi [Physics of Fluids] {10.1063/1.1761412},
  \href {https://ui.adsabs.harvard.edu/abs/1965PhFl....8.1385K} {8, 1385}

\bibitem[\protect\citeauthoryear{Kulsrud \& Pearce}{Kulsrud \&
  Pearce}{1969}]{kulsrud1969effect}
Kulsrud R.,  Pearce W.~P.,  1969, The Astrophysical Journal, 156, 445

\bibitem[\protect\citeauthoryear{Lee \& V{\"o}lk}{Lee \&
  V{\"o}lk}{1973}]{lee1973damping}
Lee M.~A.,  V{\"o}lk H.~J.,  1973, Astrophysics and Space Science, 24, 31

\bibitem[\protect\citeauthoryear{{Lee}, {Hopkins}  \& {Squire}}{{Lee}
  et~al.}{2017}]{lee:dynamics.charged.dust.gmcs}
{Lee} H.,  {Hopkins} P.~F.,   {Squire} J.,  2017, \mn@doi [\mnras]
  {10.1093/mnras/stx1097}, \href
  {http://adsabs.harvard.edu/abs/2017MNRAS.469.3532L} {469, 3532}

\bibitem[\protect\citeauthoryear{Martin, Scannapieco, Ellison, Hennawi,
  Djorgovski  \& Fournier}{Martin et~al.}{2010}]{martin2010size}
Martin C.~L.,  Scannapieco E.,  Ellison S.~L.,  Hennawi J.~F.,  Djorgovski S.,
   Fournier A.~P.,  2010, The Astrophysical Journal, 721, 174

\bibitem[\protect\citeauthoryear{{McKinnon}, {Vogelsberger}, {Torrey},
  {Marinacci}  \& {Kannan}}{{McKinnon} et~al.}{2018}]{2018MNRAS.478.2851M}
{McKinnon} R.,  {Vogelsberger} M.,  {Torrey} P.,  {Marinacci} F.,   {Kannan}
  R.,  2018, \mn@doi [\mnras] {10.1093/mnras/sty1248}, \href
  {https://ui.adsabs.harvard.edu/abs/2018MNRAS.478.2851M} {478, 2851}

\bibitem[\protect\citeauthoryear{M{\'e}nard, Scranton, Fukugita  \&
  Richards}{M{\'e}nard et~al.}{2010}]{menard2010measuring}
M{\'e}nard B.,  Scranton R.,  Fukugita M.,   Richards G.,  2010, Monthly
  Notices of the Royal Astronomical Society, 405, 1025

\bibitem[\protect\citeauthoryear{Moseley, Squire  \& Hopkins}{Moseley
  et~al.}{2019}]{moseley2019non}
Moseley E.~R.,  Squire J.,   Hopkins P.~F.,  2019, Monthly Notices of the Royal
  Astronomical Society, 489, 325

\bibitem[\protect\citeauthoryear{Oliphant}{Oliphant}{2007}]{oliphant2007python}
Oliphant T.~E.,  2007, Computing in Science \& Engineering, 9, 10

\bibitem[\protect\citeauthoryear{Pakmor, Pfrommer, Simpson  \& Springel}{Pakmor
  et~al.}{2016}]{pakmor2016galactic}
Pakmor R.,  Pfrommer C.,  Simpson C.~M.,   Springel V.,  2016, The
  Astrophysical Journal Letters, 824, L30

\bibitem[\protect\citeauthoryear{{Pan}, {Padoan}, {Scalo}, {Kritsuk}  \&
  {Norman}}{{Pan} et~al.}{2011}]{pan:2011.grain.clustering.midstokes.sims}
{Pan} L.,  {Padoan} P.,  {Scalo} J.,  {Kritsuk} A.~G.,   {Norman} M.~L.,  2011,
  \mn@doi [\apj] {10.1088/0004-637X/740/1/6}, \href
  {http://adsabs.harvard.edu/abs/2011ApJ...740....6P} {740, 6}

\bibitem[\protect\citeauthoryear{Peek, M{\'e}nard  \& Corrales}{Peek
  et~al.}{2015}]{peek2015dust}
Peek J.,  M{\'e}nard B.,   Corrales L.,  2015, The Astrophysical Journal, 813,
  7

\bibitem[\protect\citeauthoryear{Ruszkowski, Yang  \& Zweibel}{Ruszkowski
  et~al.}{2017}]{ruszkowski2017global}
Ruszkowski M.,  Yang H.-Y.~K.,   Zweibel E.,  2017, The Astrophysical Journal,
  834, 208

\bibitem[\protect\citeauthoryear{Salem, Bryan  \& Corlies}{Salem
  et~al.}{2016}]{salem2016role}
Salem M.,  Bryan G.~L.,   Corlies L.,  2016, Monthly Notices of the Royal
  Astronomical Society, 456, 582

\bibitem[\protect\citeauthoryear{{Schlickeiser}}{{Schlickeiser}}{1989}]{schlickeiser:89.cr.transport.scattering.eqns}
{Schlickeiser} R.,  1989, \mn@doi [\apj] {10.1086/167009}, \href
  {https://ui.adsabs.harvard.edu/abs/1989ApJ...336..243S} {336, 243}

\bibitem[\protect\citeauthoryear{Seligman, Hopkins  \& Squire}{Seligman
  et~al.}{2019}]{seligman2019non}
Seligman D.,  Hopkins P.~F.,   Squire J.,  2019, Monthly Notices of the Royal
  Astronomical Society, 485, 3991

\bibitem[\protect\citeauthoryear{Skilling}{Skilling}{1971}]{skilling1971cosmic}
Skilling J.,  1971, The Astrophysical Journal, 170, 265

\bibitem[\protect\citeauthoryear{{Squire} \& {Hopkins}}{{Squire} \&
  {Hopkins}}{2018}]{squire.hopkins:RDI}
{Squire} J.,  {Hopkins} P.~F.,  2018, \mn@doi [\apjl]
  {10.3847/2041-8213/aab54d}, \href
  {http://adsabs.harvard.edu/abs/2018ApJ...856L..15S} {856, L15}

\bibitem[\protect\citeauthoryear{Squire, Hopkins, Quataert  \& Kempski}{Squire
  et~al.}{2020}]{squire2020impact}
Squire J.,  Hopkins P.~F.,  Quataert E.,   Kempski P.,  2020, arXiv preprint
  arXiv:2011.02497

\bibitem[\protect\citeauthoryear{{Squire}, {Hopkins}, {Quataert}  \&
  {Kempski}}{{Squire} et~al.}{2021}]{squire:2021.dust.cr.confinement.damping}
{Squire} J.,  {Hopkins} P.~F.,  {Quataert} E.,   {Kempski} P.,  2021, \mn@doi
  [\mnras] {10.1093/mnras/stab179}, \href
  {https://ui.adsabs.harvard.edu/abs/2021MNRAS.502.2630S} {502, 2630}

\bibitem[\protect\citeauthoryear{{Strong} \& {Moskalenko}}{{Strong} \&
  {Moskalenko}}{2001}]{strong:2001.galprop}
{Strong} A.~W.,  {Moskalenko} I.~V.,  2001, \mn@doi [Advances in Space
  Research] {10.1016/S0273-1177(01)00112-0}, \href
  {https://ui.adsabs.harvard.edu/abs/2001AdSpR..27..717S} {27, 717}

\bibitem[\protect\citeauthoryear{Su et~al.,}{Su et~al.}{2020}]{su2020cosmic}
Su K.-Y.,  et~al., 2020, Monthly Notices of the Royal Astronomical Society,
  491, 1190

\bibitem[\protect\citeauthoryear{{Tielens}}{{Tielens}}{2005}]{tielens:2005.book}
{Tielens} A.~G.~G.~M.,  2005, {The Physics and Chemistry of the Interstellar
  Medium}.
Cambridge, UK: Cambridge University Press

\bibitem[\protect\citeauthoryear{Tielens, McKee, Seab  \& Hollenbach}{Tielens
  et~al.}{1994}]{tielens1994physics}
Tielens A.,  McKee C.,  Seab C.,   Hollenbach D.,  1994, Astrophys J, p.~321

\bibitem[\protect\citeauthoryear{Turk, Smith, Oishi, Skory, Skillman, Abel  \&
  Norman}{Turk et~al.}{2010}]{turk2010yt}
Turk M.~J.,  Smith B.~D.,  Oishi J.~S.,  Skory S.,  Skillman S.~W.,  Abel T.,
  Norman M.~L.,  2010, The Astrophysical Journal Supplement Series, 192, 9

\bibitem[\protect\citeauthoryear{Van Der~Walt, Colbert  \& Varoquaux}{Van
  Der~Walt et~al.}{2011}]{van2011numpy}
Van Der~Walt S.,  Colbert S.~C.,   Varoquaux G.,  2011, Computing in Science \&
  Engineering, 13, 22

\bibitem[\protect\citeauthoryear{Xu \& Yan}{Xu \& Yan}{2013}]{xu2013cosmic}
Xu S.,  Yan H.,  2013, The Astrophysical Journal, 779, 140

\bibitem[\protect\citeauthoryear{Yan \& Lazarian}{Yan \&
  Lazarian}{2002}]{yan2002scattering}
Yan H.,  Lazarian A.,  2002, Physical review letters, 89, 281102

\bibitem[\protect\citeauthoryear{{Yan} \& {Lazarian}}{{Yan} \&
  {Lazarian}}{2008}]{yan.lazarian.2008:cr.propagation.with.streaming}
{Yan} H.,  {Lazarian} A.,  2008, \mn@doi [\apj] {10.1086/524771}, \href
  {http://adsabs.harvard.edu/abs/2008ApJ...673..942Y} {673, 942}

\bibitem[\protect\citeauthoryear{{Zweibel}}{{Zweibel}}{2013}]{Zwei13}
{Zweibel} E.~G.,  2013, \mn@doi [Physics of Plasmas] {10.1063/1.4807033}, \href
  {http://adsabs.harvard.edu/abs/2013PhPl...20e5501Z} {20, 055501}

\bibitem[\protect\citeauthoryear{{van Marle}, {Casse}  \& {Marcowith}}{{van
  Marle} et~al.}{2019}]{2019MNRAS.tmp.2249V}
{van Marle} A.~J.,  {Casse} F.,   {Marcowith} A.,  2019, \mn@doi [\mnras]
  {10.1093/mnras/stz2624}, \href
  {https://ui.adsabs.harvard.edu/abs/2019MNRAS.tmp.2249V} {p.~2249}

\makeatother
\end{thebibliography}

\end{CJK}
\end{document}